\documentclass[
	aps,
	prl,
	reprint,
	amsmath,amssymb,
	superscriptaddress
]{revtex4-2}

\usepackage{graphicx}
\usepackage{dcolumn}
\usepackage{bm}

\usepackage{color,soul} 
\usepackage{comment} 	
\usepackage{xcolor} 

\usepackage{verbatim} 

\usepackage[bookmarks]{hyperref} 	

\setul{0.3ex}{0.25ex} 
\setulcolor{magenta}

\begin{document}
\title{Persistent response in ultra-strongly driven mechanical membrane resonators}

\date{\today}

\author{Fan Yang}
\affiliation{Fachbereich Physik, Universit{\"a}t Konstanz, 78457 Konstanz, Germany}
\author{Felicitas Hellbach}
\affiliation{Fachbereich Physik, Universit{\"a}t Konstanz, 78457 Konstanz, Germany}
 \author{Felix Rochau}
 \affiliation{Fachbereich Physik, Universit{\"a}t Konstanz, 78457 Konstanz, Germany}
 %
 %
 \author{Wolfgang Belzig}
 \affiliation{Fachbereich Physik, Universit{\"a}t Konstanz, 78457 Konstanz, Germany}
\author{Eva M. Weig}
\affiliation{Fachbereich Physik, Universit{\"a}t Konstanz, 78457 Konstanz, Germany}
\affiliation{Fakultät für Elektrotechnik und Informationstechnik, Technische Universit{\"a}t M{\"u}nchen, 80333 München, Germany}

\author{Gianluca Rastelli}
\email{gianluca.rastelli@ino.cnr.it}
\affiliation{Fachbereich Physik, Universit{\"a}t Konstanz, 78457 Konstanz, Germany}
\affiliation{INO-CNR BEC Center and Dipartimento di Fisica, Universit{\`a}  di Trento, 38123 Povo, Italy}

\author{Elke Scheer}%
 \email{elke.scheer@uni-konstanz.de}
\affiliation{Fachbereich Physik, Universit{\"a}t Konstanz, 78457 Konstanz, Germany}%
\begin{abstract}
We study experimentally and theoretically the phenomenon of persistent response in ultra-strongly driven membrane resonators. 
This term denotes the development of a vibrating state with nearly constant amplitude over an extreme wide frequency range. 
We reveal the underlying mechanism of the persistent response state by directly imaging the vibrational state using advanced optical interferometry. 
We argue that the persistent state is related to the nonlinear interaction between higher order flexural modes and higher-order overtones of the driven mode. 
Finally, we propose a stability diagram for the different vibrational states that the membrane can adopt.
\end{abstract}

\maketitle
%
%
%
%
%
%
Mechanical resonators have applications as ultra-sensitive sensors of e.g. molecular transport \cite{arash2015review} 
or as nanomechanical logic gates \cite{mahboob2008bit, guerra2010noise, tadokoro2021highly}. 
Nonlinear mechanical properties have come into focus for signal enhancement \cite{rugar1991mechanical, karabalin2011signal,papariello2016ultrasensitive} 
and noise reduction in metrology \cite{chowdhury2020weak,huber2020spectral} for signal processing. 
In membrane resonators, different flexural modes can be excited and have been utilized e.g. for coupling mechanical energy to other degrees of freedom such as light and atoms \cite{camerer2011realization, purdy2013observation, andrews2014bidirectional, jockel2015sympathetic,xu2016topological,karg2020light}. 
For micro- and nano-mechanical systems in the nonlinear regime the Duffing model is widely used to describe the vibrational behavior \cite{nayfeh2008nonlinear,schuster2009reviews}.
The nonlinearity can be caused by motion-induced tension, i.e., the oscillation of one mode induces an elongation of the resonator that affects the dynamics also of another mode. Hence, for large deflection amplitudes, displacement-induced changes of the mechanical properties may give rise to nonlinear mode coupling \cite{schuster2009reviews, westra2010nonlinear}. 
Membrane oscillators in the strongly nonlinear regime offer the possibility to explore nonlinear mode coupling mechanisms \cite{nayfeh2008nonlinear,manevich2005mechanics,mangussi2016internal,shoshani2017anomalous, guttinger2017energy, chen2017direct, yang2019spatial}.
Such coupling leads to exotic line shapes of the response function due to the transfer of energy between mechanical modes that can be faster than the energy relaxation time \cite{guttinger2017energy,chen2017direct,yang2019spatial}.

\begin{figure}[t]
   \includegraphics[width=\linewidth]{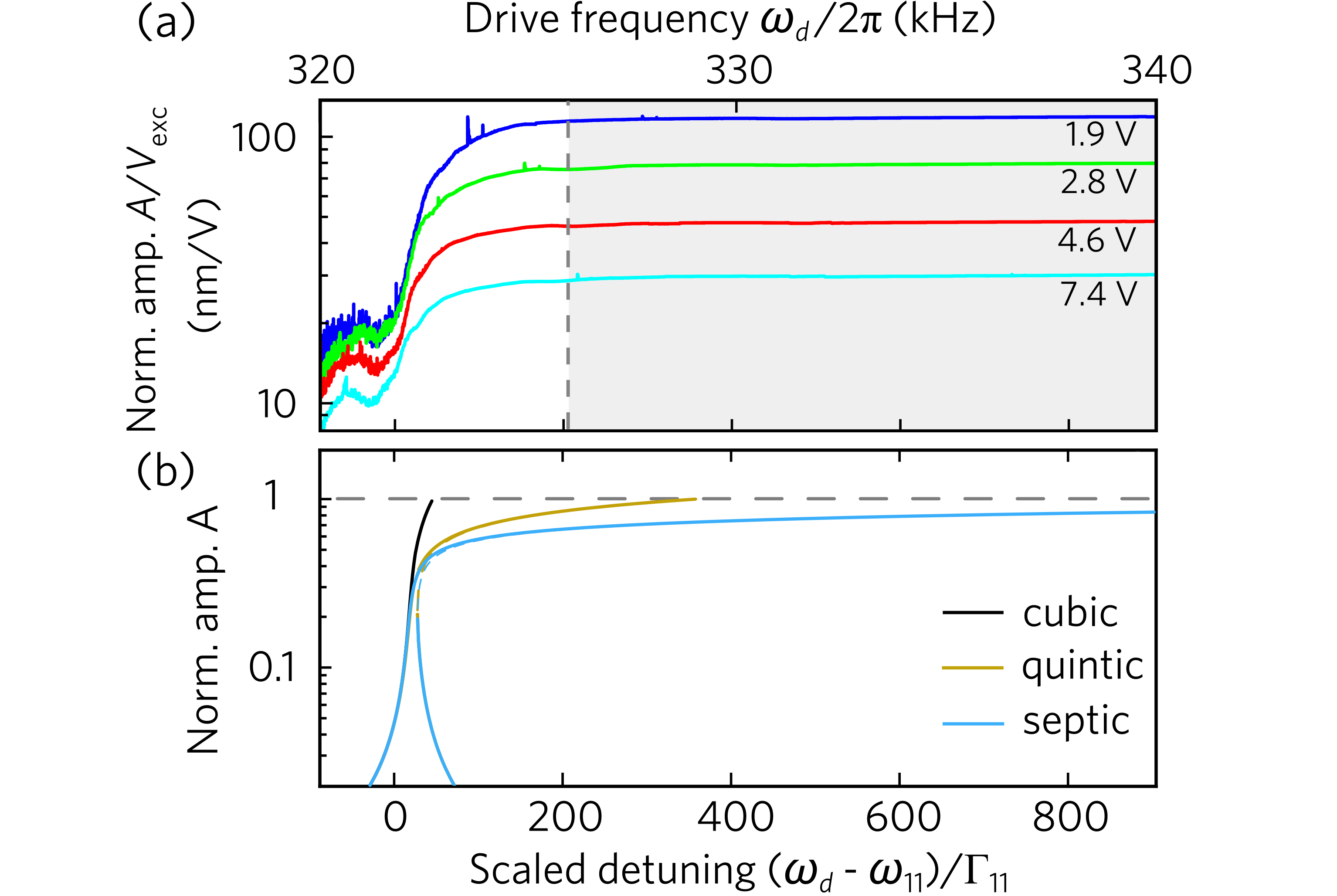}
   \caption{
   Persistent response and higher-order nonlinearities. 
       (a) Response function generated by different drive voltages recorded by imaging white light interferometrey, showing the mean amplitude response        (normalized to the excitation voltage and averaged over the whole membrane area) of (1,1) mode with eigen-frequency around 321 kHz. 
       (b) Theoretical  curves for the amplitude of the (1,1) mode with different higher-order nonlinear forces.  
    }
   \label{fig:persistent_curve}
\end{figure}

%
%
%
%
Here we report a previously undescribed nonlinear behavior achieved in ultra-strongly driven resonators, which occurs when we apply a single sinusoidal drive at a frequency in the range of a flexural mode of a membrane resonator. For simplicity we concentrate here on the fundamental mode ((1,1) mode); similar observations have been made for other flexural modes, though.  
Varying the drive frequency, we observe the development of a plateau in the response curves that extends over a considerable frequency range with a nearly constant amplitude, see Fig.~\ref{fig:persistent_curve}(a). %
We denote the almost constant amplitude over an extremely wide frequency range as persistent response. 
We argue that this state is maintained by two different interaction mechanisms: nonlinear coupling between different flexural modes and spatial modulation of overtones \cite{yang2019spatial} (i.e. the appearance of frequency multiples of one flexural mode). 
Here we focus on the latter mechanism which becomes progressively dominant at large detuning.
In the final part, we make a connection with our previous results \cite{yang2019spatial} and propose a stability diagram for the vibrational state of the membrane under ultra-strong drive.

%
%
%
%
%
%
%

%
%
%
%
The sample fabrication and measurement principles have been described in detail elsewhere \cite{waitz2012mode,yang2017quantitative,yang2019spatial}. 
In brief, a chip carrying a high-stress silicon nitride membrane is glued onto a piezo ring and installed in a vacuum chamber. 
Vibrations of the membrane are excited by applying an AC voltage $V_\textrm{exc} \cdot \sin (\omega_dt$) to the piezo resulting in an inertial excitation of the membrane. 
The vibrational states of a specific $(m,n)$ mode (the integers $m$ and $n$ indicating the number of deflection nodes in $x$- and $y$-direction) of the membrane are observed by imaging white light interferometry (IWLI), 
spatially resolving the deflection profile and obtaining the averaged mean amplitude response over the membrane surface area \cite{petitgrand20013d,yang2019spatial}, 
and on the other hand by Michelson interferometry (MI) focusing on one particular position of the membrane with a spot diameter of $\mathrm{\sim} 1\,\mu$m. 
Further experimental details are given in the Supplemental Material (SM) \cite{SM}.

%
%
%
%
We utilize the IWLI signal integrated over the entire membrane area to record the nonlinear vibration behavior under ultra-strong excitation 
and with a driving frequency $\omega_d$ around $\omega_{11}$. 
Four selected traces revealing the pronounced amplitude saturation recorded for different $V_\textrm{exc}$ are plotted over a limited frequency range in Fig. \ref{fig:persistent_curve}(a). 
The complete frequency-vs-drive map is shown in Fig. S3 in the SM \cite{SM}. 
The flattening of the response curve for small detuning above the linear eigenfrequency  (here up to ~324 kHz) can be described by the spatial modulation of localized overtones as explained in Ref. \cite{yang2019spatial}. 
We here concentrate on the frequency range in which the amplitude is almost independent of the drive frequency  highlighted by the grey area in Fig. \ref{fig:persistent_curve}(a). At even larger detuning, we observe features that deviate from the smooth curve of the persistent response, see below.

By expanding the complex elastic energy potential of the membrane in terms of the normal modes, one generates all possible nonlinear interactions between the different modes.
The generic term reads
\begin{align}
V_{ (n\,m \, |\,  \ell\,p) }^{(kh)}
\,\,\,\, = \,\,\,\,
\lambda_{ (n\,m \, |\,  \ell\,p) }^{(kh)} \,\,\, q_{n\, m}^k  \,\,\,\,  q_{\ell\, p}^h \, , 
\end{align}
with the coupling strength $\lambda_{ (n\,m \, |\,  \ell\,p) }^{(kh)}$ where  $k,h$ are integers and $q_{n\, m}$ and $q_{\ell\, p}$ the amplitudes of the two modes $(n,m)$ and $(\ell,p)$.
Except for some frequency ranges, where we observe features that deviate from the smooth curve of the persistent response,
we assume the higher-order modes and/or the overtones are weakly excited due to nonlinear interaction with the fundamental mode $(1,1)$ such that they oscillate in their linear or Duffing state.\\
\indent  However, the dynamics of the higher-order modes affects the response of the driven mode.
To illustrate this idea, we discuss a model in which we consider the mode $(1,1)$ coupled to the three modes $(n,n)$ with $n=2,3,4$ 
through the potential $\sum_{n=2}^{4} V_{ (1\,1 \, |\,  n\,n) }^{(n1)} = \sum_{n=2}^{4}  \lambda_{ (1\,1 \, |\,  n\,n) }^{(n1)} \,\,\, q_{1\, 1}^n  q_{n\, n}$.
If the higher-order modes are in the harmonic regime, as shown in \cite{SM}, 
the dynamic of the fundamental mode driven by a linear force can be described by effective high-order nonlinearities as follows
\begin{align}
\label{Eq_nonlin_fund}
\ddot{q}_{11}(t)=   
&-\omega_{11}^2q_{11}(t)-2\Gamma_{11}\dot{q}_{11}(t)+F \cos(\omega_d t)-\gamma_1 q^3_{11}(t) \nonumber \\
&- \mu q^5_{11}(t)- \nu q^7_{11}(t) \, ,
\end{align}
where $\gamma$ is the Duffing nonlinearity whereas  the quintic and the septic nonlinearity have coefficients $\mu>0$ and $\nu>0$. 
Using the rotating wave approximation (RWA) \cite{SM}, 
we determine the maximum amplitude and detuning 
\footnote{
Note that the inclusion of higher-order terms causes still three possible solutions in the parameter regime of the experiment, two stable and an unstable one, as in the Duffing case,
not shown in in Fig. \ref{fig:persistent_curve}(b).
}.
For a sufficiently strong drive, the vibration amplitude is large and the nonlinear higher-order terms become increasingly important. 
Figure \ref{fig:persistent_curve}(b) displays that the curve  progressively flattens by adding nonlinear self-interaction terms. 
Furthermore setting $A_{max}$ as the maximum achievable amplitude at a given drive force, we have that
\begin{equation}
\left( \omega_{d,max} - \omega_{11} \right)   = \frac{3\gamma_1}{8\omega_{11}} A_{max}^2 +  \frac{5\mu}{16\omega_{11}} A_{max}^4 +  \frac{35\nu}{128\omega_{11}} A_{max}^6 \, .
\end{equation}

\begin{figure}[t]
 \includegraphics[width=\linewidth]{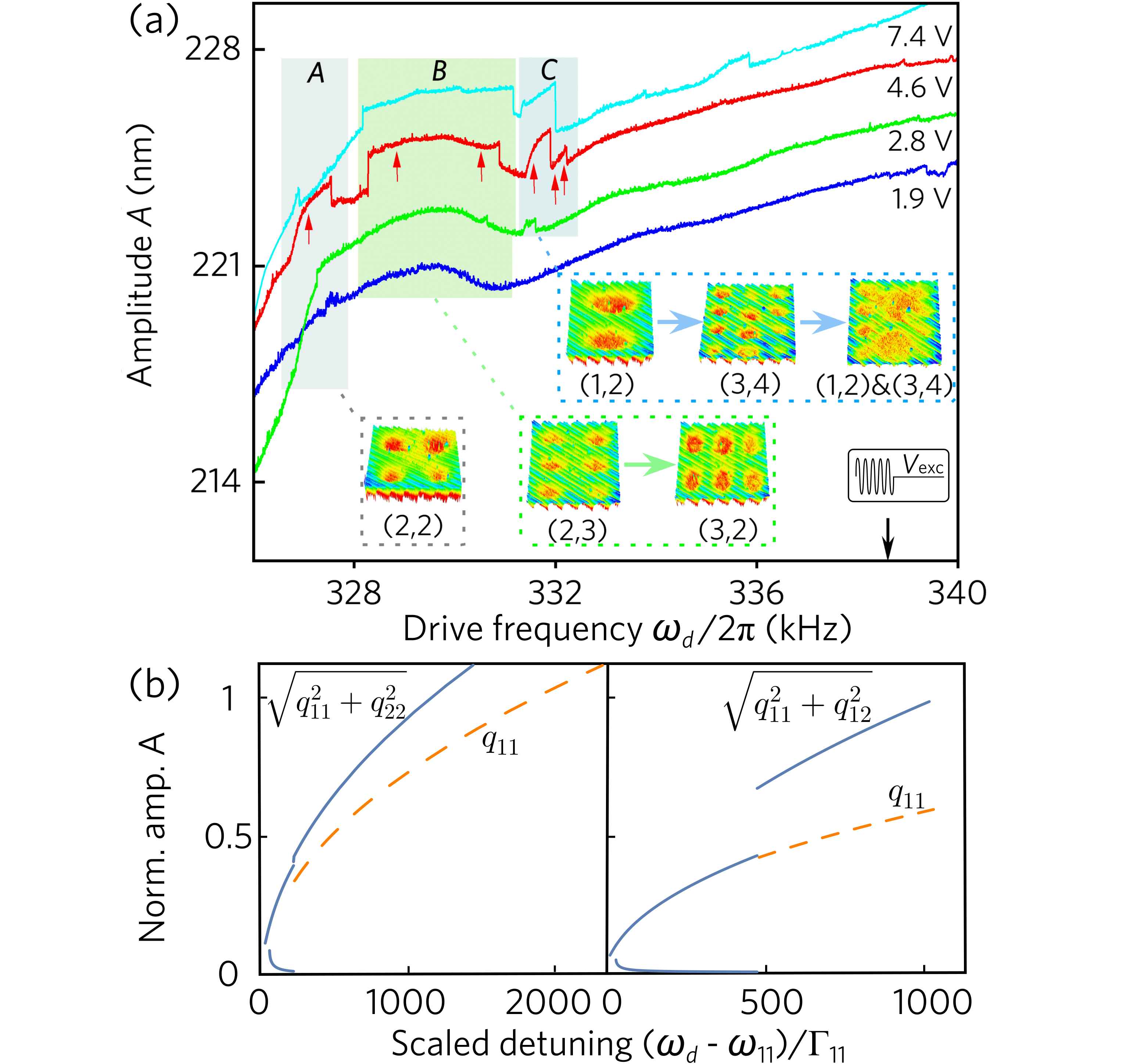}
 \caption{Persistent response and visualization of deflection patterns. 
 (a) Blow-up of the grey area of Fig. \ref{fig:persistent_curve}(a) without normalizing the y-axis. 
 The plateau reveals small steps and kinks superimposed, some of them being marked by colored areas \textit{A}, \textit{B} and \textit{C}. 
 The red arrows indicate the positions where we captured the deflection patterns for $V_\textrm{exc}$ = 4.6 V shown in the insets. 
 The black arrow indicates the frequency at which the ring-down experiments shown in Fig. \ref{fig:rd_and_phase}(a) are performed. 
 (b) Numerical solutions of the nonlinear coupling model (solid lines: coupled modes, dashed lines: driven modes).  
  Left: model interaction between the mode $(1,1)$ and the mode $(2,2)$; 
  Right: model interaction between the mode $(1,1)$ and the mode $(1,2)$ (see text for details).
  }
   \label{fig:persistent_zoom}
\end{figure}

%
%
%
%
Figure \ref{fig:persistent_zoom}(a) shows a zoom into the grey-shaded area in Fig. \ref{fig:persistent_curve}(a), where the amplitude is almost independent of the drive $V_\textrm{exc}$. 
However, nanometer scale variations are detected in all four curves, signaled by small steps and kinks on the plateaus. 
%
%
%
%
%
%
%
%
%
%
%
%
%
%
For $\omega_d ~\mathrm{<} ~\omega_{22}$/2 we only see the deflection pattern of the (1,1) mode. 
When $\omega_d ~\mathrm{\approx} ~\omega_{22}$/2, a (2,2) mode pattern is observed superimposed over the (1,1) mode pattern in the area \textit{A}. 
Further increasing $\omega_d$, the (2,3) mode is switched on abruptly and then transitions into the (3,2) mode in the region \textit{B}. 
For even larger $\omega_d$, the pattern of the (1,2) mode appears which then switches to (3,4) pattern at the first shoulder in area \textit{C}.
Before they disappear again, both patterns are observed simultaneously.

The appearance of higher order modes $(m,n)$ in the deflection profile, is a signature of nonlinear mode coupling.
Together with the possibility to excite overtones of the individual ($m$,$n$) modes \cite{yang2019spatial}, we argue that the complex superposition of several modes 
is generated by an effective nonlinear coupling between different types of modes and overtones.
The appearance of changes of the slope was explained in \cite{yang2019spatial} in terms of standard nonlinear resonance, e.g. $1:k$ coupling. 
Here we discuss the discontinuous steps.
They can be explained by an \textit{indirect parametric nonlinear interaction} mediated by the overtones of the fundamental mode ($m = n$) or \textit{direct parametric nonlinear interaction} between the modes ($m \neq n$).
We discuss two examples to illustrate this idea.

The first example is the indirect interaction of the driven mode $(1,1)$ with the $(2,2)$ mode, 
with eigenfrequency $\omega_{22} \simeq 2\,\omega_{11}$.
We consider the nonlinear interaction term
\begin{align}
\label{eq:int_1}
V_{ (1\,1 \, |\,  2\,2) }^{(22)} \,\, = \,\,  \frac{1}{2} \,\, \lambda_{ (1\,1 \, |\,  2\,2) }^{(22)}  \,\, q_{11}^2 \,\, q_{22}^2
\end{align}
between the two Duffing resonators $q_{11}$ and $q_{22}$ (for simplicity and qualitative analysis, we neglect high order nonlinearities of the fundamental mode). 
Due to the presence of an overtone of the fundamental mode, we use the 
ansatz $q_{11} \simeq \left( u_{1} e^{i\omega_d t}  + u_{3} e^{i3 \omega_d t} +\mbox{c.c.} \right) /2$.
\footnote{The first overtone at $2\,\omega_d$ is also present and couples to another higher-order mode, see Supplemental Material \cite{SM} for further examples.} 
Then, the mixing term $q_{11}^2 \propto  u_{1}  u_{3} e^{i4 \omega_d t} $  in Eq.~(\ref{eq:int_1}) represents the parametric excitation for the high frequency mode $(2,2)$,
which becomes resonant and relevant  as soon as $4\,\omega_d \approx  2\,\omega_{22}$. 
Using the RWA (see SM \cite{SM}), we obtain the numerical solutions for the amplitudes of the $(1,1)$ and the $(2,2)$ mode shown in the left panel of Fig. \ref{fig:persistent_zoom}(b). 

%
%
%
%
 \begin{figure}[t]
   \includegraphics[width=\linewidth]{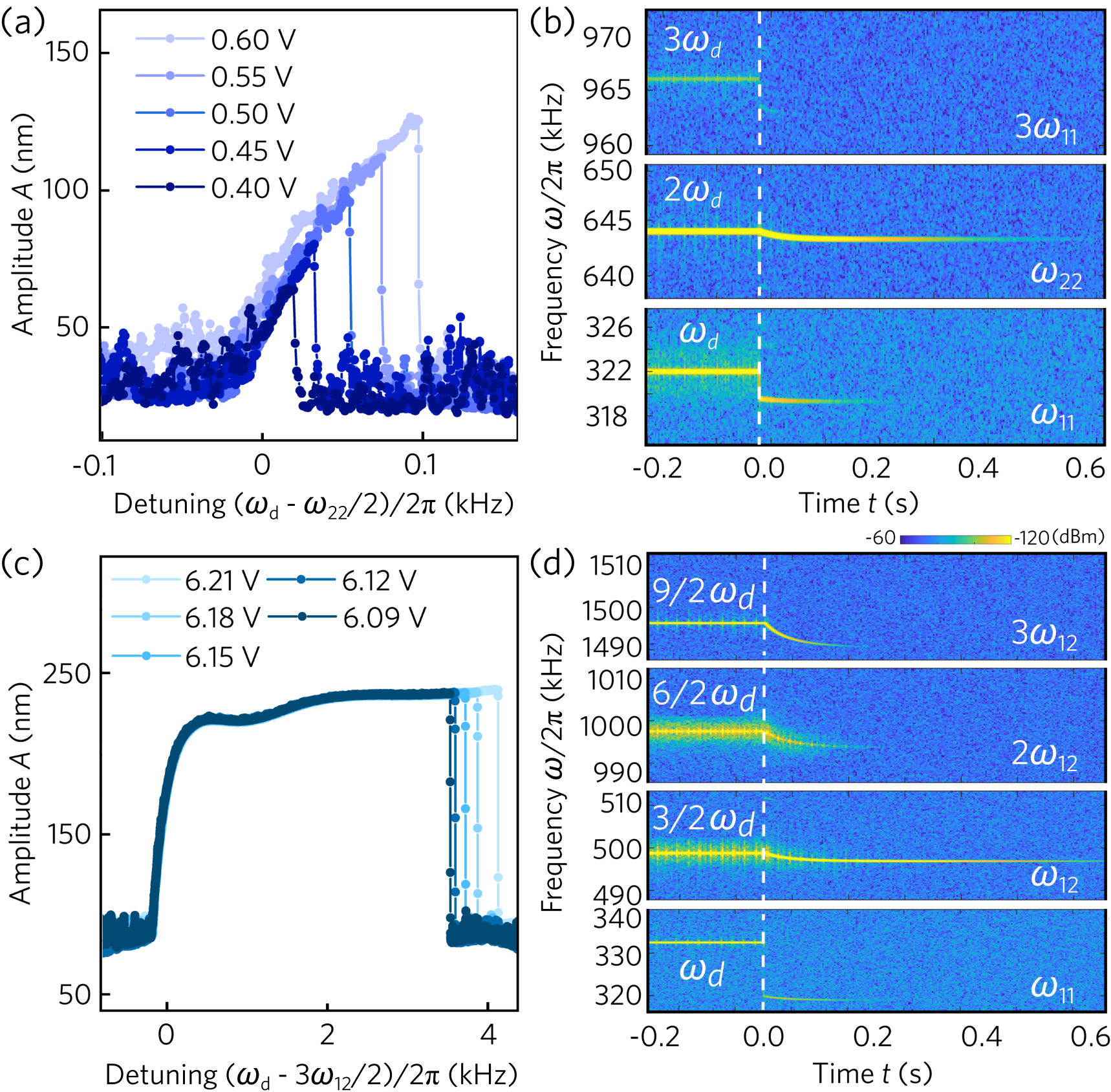}
   \caption[width=\textwidth]{
   The subharmonic response functions of the flexural mode $(2,2)$ [(a),(b)] and the mode $(1,2)$ [(c),(d)] 
       (a) Subharmonic amplitude response of the (2,2) mode. 
    (b) Frequency spectrum of the ring-down for $V_\textrm{exc}\,=\,5.0\,\textrm{V}$ and $\omega_d/2\pi$ = 322.0 kHz (instantly, $\omega_{11}/2\pi = 319\,\textrm{kHz}$) showing three integer overtones of $\omega_d$.
    The drive is turned off at $t = 0$. 
    (c) Subharmonic response of the flexural mode $(1,2)$. 
    (d) Same as (b) but for $V_\textrm{exc}$ = 7.0 V and $\omega_d/2\pi = 332.5~\textrm{kHz}$, showing three half-integer overtones of $\omega_d$. 
    The energy decay during ring-down is shown in the SM \cite{SM}.
    }
   \label{fig:sub_driven_22_12}
 \end{figure}
%
%
%
%

As a second example, we consider the direct interaction between the driven mode $(1,1)$ and the $(1,2)$ mode with eigenfrequency $\omega_{12}\approx \frac{3}{2} \,\omega_{11}$. 
These modes are described as before as two Duffing resonators for which we have the nonlinear potential 
\begin{align}
\label{eq:int_2}
V_{ (1\,1 \, |\,  1\,2) }^{(32)} \,\, = \,\,  \frac{1}{2} \,\, \lambda_{ (1\,1 \, |\,  1\,2) }^{(32)}  \,\, q_{11}^3 \,\, q_{12}^2 .
\end{align}
This parametric interaction becomes resonant and relevant when $3\,\omega_d \approx 2\,\omega_{12}$: In the nonlinear interaction Eq.~(\ref{eq:int_2}) 
the term $q_{11}^3 \propto u_1^3 e^{i 3 \omega_d t}$ represents the parametric excitation for the mode $(1,2)$.
For this second example, the numerical solution of the amplitudes is shown in the right panel of Fig. \ref{fig:persistent_zoom}(b). 
 
To obtain further insight into this nonlinear interaction, $\omega_d$ is swept upward from a starting frequency above the linear eigenfrequency (here $\omega_d/2\pi > 326 $kHz) such that the mode $(1,1)$ vibrates in its low-amplitude state (amplitude $<$ 50\,nm).
The results are shown in  Fig. \ref{fig:sub_driven_22_12}. 
For the first example, the curves in Fig. \ref{fig:sub_driven_22_12}(a) show the subharmonic amplitude response of mode $(2,2)$ 
for different excitation strengths. 
Figure \ref{fig:sub_driven_22_12}(b) displays a series of power spectra taken during the ring-down of the mode. 
As soon as the drive has been switched off,
the components $\omega_d$  and  $3\,\omega_d$  show a jump which is typical for a nonlinear vibration in low amplitude (harmonic) state.
These components can be associated with the fundamental mode with $\omega_{11}$ and its odd overtone $3\,\omega_{11}$ ($\neq \omega_{33}$). 
On the other side, the components at $2\,\omega_d$ of the mode $(2,2)$ decays smoothly.
Notice that the majority of the power is observed around $\omega_{22}$. 
All these features are consistent with mode $(2,2)$ being parametrically driven by the mode $(1,1)$ and its odd overtone due to nonlinear coupling according to Eq.~(\ref{eq:int_1}).

In Fig. \ref{fig:sub_driven_22_12}(c) we show the subharmonic response of the mode  $(1,2)$, when the fundamental mode oscillates in the low-amplitude state. 
Upon increasing $\omega_d$, the amplitude is suddenly pumped up when $V_\textrm{exc} >6.09~\textrm{V}$. 
This is characteristic for parametric coupling.
Here, the parametric excitation is stronger than in the previous example since the nonlinear interaction with the mode $(1,1)$ in Eq.~(\ref{eq:int_2}) is direct and not mediated by overtones.
The ring-down in Fig. \ref{fig:sub_driven_22_12}(d) is performed for $\omega_d /2\pi=\frac{2}{3}\,\omega_{12}/2\pi = 332.5~\textrm{kHz} $ and $V_\textrm{exc} = 7~\textrm{V}$. 
The corresponding power spectra reveal the half-integer frequencies $\frac{3}{2}\,\omega_d$, $\frac{6}{2}\,\omega_d$ and $\frac{9}{2}\,\omega_d$ 
which decay to $\omega_{12}$, 2$\,\omega_{12}$ and 3$\,\omega_{12}$ after the drive is switched off.
Notice the absence of the overtones of the mode $(1,1)$.
Again,  these features are consistent with a parametrical drive of the mode $(1,2)$ by the mode $(1,1)$.

As an additional example, we shortly discuss the possible excitation mechanism of the pattern observed in area \textit{B} of Fig. \ref{fig:persistent_zoom}(a): The eigenfrequencies $\omega_{23}$ and $\omega_{32}$ differ slightly from each other, because of the small deviation of the membrane from a square shape. 
The 2\textsuperscript{nd} overtones of these frequencies, 2$\,\omega_{23}/2\pi = 1636\,\textrm{kHz} $ and $2\,\omega_{32}/2\pi = 1624\,\textrm{kHz}$ are close to 5$\,\omega_d$. 
We therefore argue that these modes are excited subharmonically via nonlinear interaction with the fundamental mode of type $\propto q_{11}^5 q_{23}^2$ or $\propto q_{11}^5 q_{32}^2$.

We will now turn to the analysis of the persistent response of the $(1,1)$ mode. 
%
%
%
%
%
%
%
%
\begin{figure}[t]
   \begin{center}
    \includegraphics[width=\linewidth]{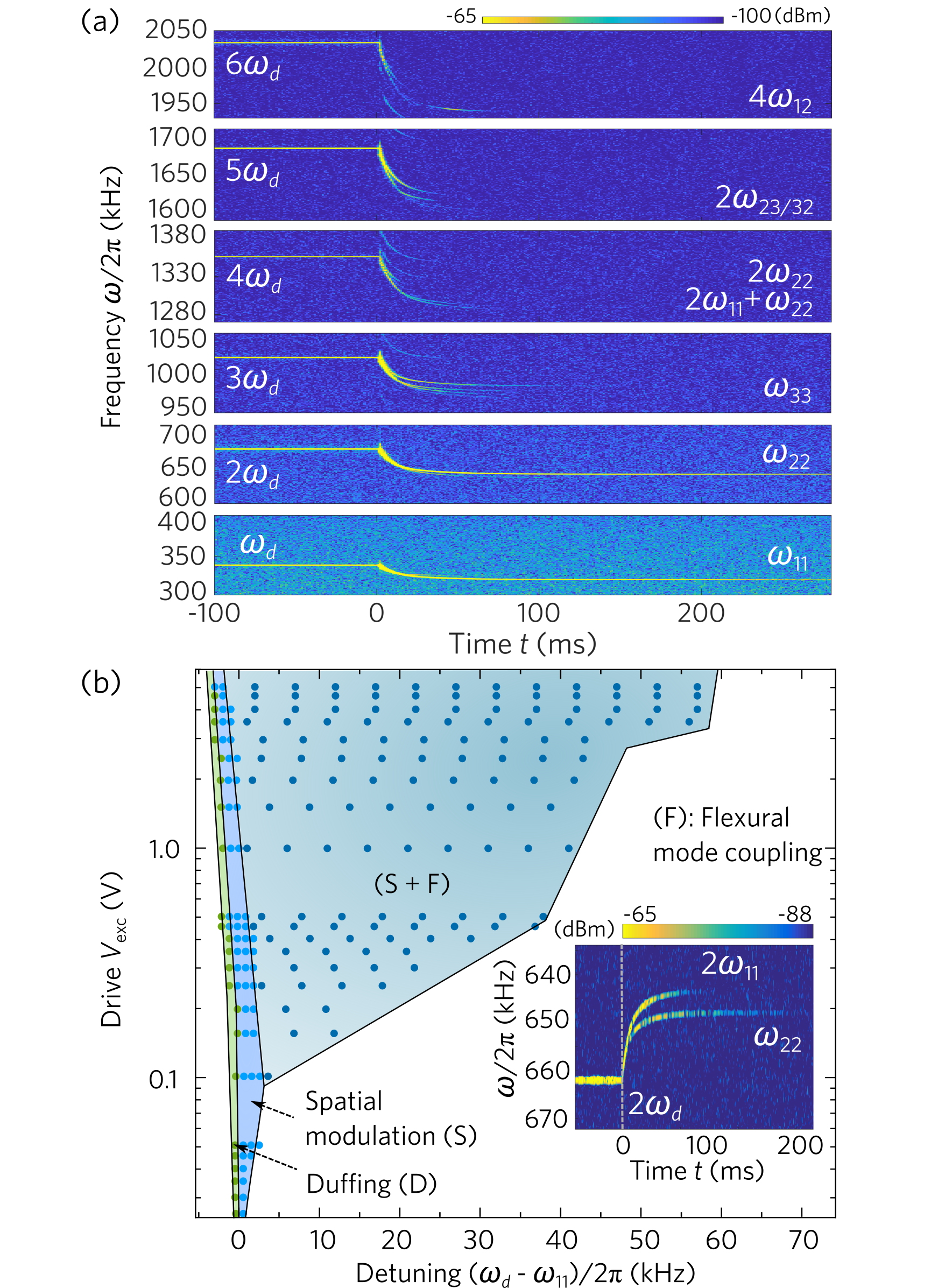}
   \caption[width=\textwidth]{
    Ring-down measurement in the flexural mode coupling regime. 
    (a) Frequency spectrum of the ring-down for $V_\textrm{exc}$ = 5 V and $\omega_d$ = 338.6 kHz showing six overtones of $\omega_d$. 
    The drive is turned off at $t = 0$. 
    For details of the mode assignment and energy decay see SM \cite{SM}. 
    (b) Stability diagram of the vibrational state of the membrane resonator in the strong nonlinear response as a function of detuning from $\omega_\textrm{(1,1)}$, and drive $V_\textrm{exc}$. 
    The solid lines are determined by combining experimental data from MI (colored dots) and IWLI data shown in Fig. S3 in the SM \cite{SM}. 
    The inset shows an example of a coexistence of flexural mode coupling and spatial overtones \cite{yang2019spatial}:
     Ring-down recorded for $V_\textrm{exc} = 1.77~\textrm{V}$ at $\omega_d/2\pi = 331~\textrm{kHz} $ showing a frequency splitting into $\omega_{22}$ and 2$\,\omega_{11}$.
     }
   \label{fig:rd_and_phase}
    \end{center}
\end{figure}
%
%
%
%
%
%
%
We repeat the experiment shown in Fig. \ref{fig:persistent_curve}(a) 
to explore the frequency response and the ring-down behavior of the membrane subject to an ultra-strong excitation, $V_\textrm{exc}$ = 5.0 V at $\omega_d/2\pi$ = 338.6 kHz, 
marked as black arrow in Fig. \ref{fig:persistent_zoom}(a), well above $\omega_{11}$ using MI. 
With the excitation switched off (t = 0), the time-resolved power spectrum recorded in the frequency range 250 kHz to 2.5 MHz displayed 
in Fig. \ref{fig:rd_and_phase}(a) shows the contribution of $\omega_d$ and its overtones up to 6\textsuperscript{th} order. 
Owing to the detuned drive $(n\,\omega_d \neq \omega_{nn})$, 
the frequency of the individual contributions shifts towards a nearby eigenfrequency of the membrane \cite{antoni2013nonlinear, polunin2016characterization}. 
By comparison with the eigenfrequencies, the most prominent modes are identified as
(i) flexural modes with $m = n$ and their overtones (e.g.  the 2\textsuperscript{nd} overtones of the modes $(1,1)$ and $(2,2)$);
(ii) flexural modes with $m \neq n$, here the 2\textsuperscript{nd} overtone of the (2,3) and the (3,2) modes and the 4\textsuperscript{th} overtone of the (1,2) mode;
(iii) mixed frequency response such as 2$\,\omega_{11}$ + $\omega_{22}$ in the decay of 4$\,\omega_d$ (the detailed identification of the mixed-frequency modes is shown in the SM \cite{SM}).

Finally, combining the findings in the various regimes enables us to establish a tentative stability diagram of the complex vibrational behavior of the persistent response of the membrane resonator subject to strong excitation, as shown in Fig. \ref{fig:rd_and_phase}(b). 
When increasing the excitation strength beyond the linear response regime, a gradual transition to the usual Duffing behavior occurs (green shaded area). 
For even larger $V_\textrm{exc}$, the spatial modulation regime with overtones sets in in the blue-shaded area labeled S. 
When increasing $\omega_d$, this state transitions gradually to the flexural mode coupling state. 
In the transition range (green shaded area labeled S+F),  flexural modes and spatially modulated overtones of the fundamental mode are observed simultaneously, as exemplified in the inset. 
Here, a ring-down experiment is shown for the frequency range around $\omega_d$.
After turning off the drive, the components at $2\,\omega_d$ splits into two branches that decay into the 2$\,\omega_{11}$ and $\omega_{22}$ frequencies, namely to one higher flexural mode and to one spatial overtone of the fundamental mode. 

Summarizing, by combination of two complementary experimental approaches and with nonlinear coupling models, 
we described the persistent response effect of the membrane resonator and 
revealed unconventional nonlinear interaction between the driven fundamental mode, its overtones and higher-order flexural modes, which occurs under ultra-strong drive. 
This persistent response state has, in the example shown here, an extension of more than 50\%  of the eigenfrequency of the driven mode, but even longer plateaus are shown in the SM \cite{SM}. 
Hence, ultra-strongly driven membranes lend themselves to be utilized as adaptive resonators without actively tuning their eigenfrequency, e.g. for energy transfer to resonating systems with different degrees of freedom.
So far, we have investigated membranes with quality factors $Q$ around 20,000. 
We argue that the observed phenomena might become even more pronounced when increasing the $Q$ factor by using thinner membranes. 
We expect that our findings will help clarifying other unusual phenomena appearing in strongly nonlinear systems, also beyond the mechanical case. 

\acknowledgments
The authors thank J. S. Ochs and A. Brieussel for their contributions in the early stage of the work. 
We are indebted to T. Dekorsy, V. Gusev, M. Hettich, P. Leiderer, M. Fu and Y. Jiang for fruitful discussion and comments about the work. 
The authors gratefully acknowledge financial support from the China Scholarship Council, 
the European Union's Horizon 2020 program for Research and Innovation under grant agreement No. 732894 (FET Proactive HOT), 
the Deutsche Forschungsgemeinschaft  (DFG, German Research Foundation) through Project-ID 32152442 - SFB 767 and 
Project-ID 425217212 - SFB 1432, and from the German Excellence Strategy via the Zukunftskolleg of the University of Konstanz.

\bibliographystyle{apsrev4-1}

\bibliography{references}

\begin{thebibliography}{32}%
\makeatletter
\providecommand \@ifxundefined [1]{%
 \@ifx{#1\undefined}
}%
\providecommand \@ifnum [1]{%
 \ifnum #1\expandafter \@firstoftwo
 \else \expandafter \@secondoftwo
 \fi
}%
\providecommand \@ifx [1]{%
 \ifx #1\expandafter \@firstoftwo
 \else \expandafter \@secondoftwo
 \fi
}%
\providecommand \natexlab [1]{#1}%
\providecommand \enquote  [1]{``#1''}%
\providecommand \bibnamefont  [1]{#1}%
\providecommand \bibfnamefont [1]{#1}%
\providecommand \citenamefont [1]{#1}%
\providecommand \href@noop [0]{\@secondoftwo}%
\providecommand \href [0]{\begingroup \@sanitize@url \@href}%
\providecommand \@href[1]{\@@startlink{#1}\@@href}%
\providecommand \@@href[1]{\endgroup#1\@@endlink}%
\providecommand \@sanitize@url [0]{\catcode `\\12\catcode `\$12\catcode
  `\&12\catcode `\#12\catcode `\^12\catcode `\_12\catcode `\%12\relax}%
\providecommand \@@startlink[1]{}%
\providecommand \@@endlink[0]{}%
\providecommand \url  [0]{\begingroup\@sanitize@url \@url }%
\providecommand \@url [1]{\endgroup\@href {#1}{\urlprefix }}%
\providecommand \urlprefix  [0]{URL }%
\providecommand \Eprint [0]{\href }%
\providecommand \doibase [0]{http://dx.doi.org/}%
\providecommand \selectlanguage [0]{\@gobble}%
\providecommand \bibinfo  [0]{\@secondoftwo}%
\providecommand \bibfield  [0]{\@secondoftwo}%
\providecommand \translation [1]{[#1]}%
\providecommand \BibitemOpen [0]{}%
\providecommand \bibitemStop [0]{}%
\providecommand \bibitemNoStop [0]{.\EOS\space}%
\providecommand \EOS [0]{\spacefactor3000\relax}%
\providecommand \BibitemShut  [1]{\csname bibitem#1\endcsname}%
\let\auto@bib@innerbib\@empty
\bibitem [{\citenamefont {Arash}\ \emph {et~al.}(2015)\citenamefont {Arash},
  \citenamefont {Jiang},\ and\ \citenamefont {Rabczuk}}]{arash2015review}%
  \BibitemOpen
  \bibfield  {author} {\bibinfo {author} {\bibfnamefont {B.}~\bibnamefont
  {Arash}}, \bibinfo {author} {\bibfnamefont {J.-W.}\ \bibnamefont {Jiang}}, \
  and\ \bibinfo {author} {\bibfnamefont {T.}~\bibnamefont {Rabczuk}},\
  }\bibfield  {title} {\emph {\bibinfo {title} {A review on nanomechanical
  resonators and their applications in sensors and molecular transportation},\
  }}\href@noop {} {\bibfield  {journal} {\bibinfo  {journal} {Appl. Phys.
  Rev.}\ }\textbf {\bibinfo {volume} {2}},\ \bibinfo {pages} {021301} (\bibinfo
  {year} {2015})}\BibitemShut {NoStop}%
\bibitem [{\citenamefont {Mahboob}\ and\ \citenamefont
  {Yamaguchi}(2008)}]{mahboob2008bit}%
  \BibitemOpen
  \bibfield  {author} {\bibinfo {author} {\bibfnamefont {I.}~\bibnamefont
  {Mahboob}}\ and\ \bibinfo {author} {\bibfnamefont {H.}~\bibnamefont
  {Yamaguchi}},\ }\bibfield  {title} {\emph {\bibinfo {title} {Bit storage and
  bit flip operations in an electromechanical oscillator},\ }}\href@noop {}
  {\bibfield  {journal} {\bibinfo  {journal} {Nat. Nanotechnol.}\ }\textbf
  {\bibinfo {volume} {3}},\ \bibinfo {pages} {275} (\bibinfo {year}
  {2008})}\BibitemShut {NoStop}%
\bibitem [{\citenamefont {Guerra}\ \emph {et~al.}(2010)\citenamefont {Guerra},
  \citenamefont {Bulsara}, \citenamefont {Ditto}, \citenamefont {Sinha},
  \citenamefont {Murali},\ and\ \citenamefont {Mohanty}}]{guerra2010noise}%
  \BibitemOpen
  \bibfield  {author} {\bibinfo {author} {\bibfnamefont {D.~N.}\ \bibnamefont
  {Guerra}}, \bibinfo {author} {\bibfnamefont {A.~R.}\ \bibnamefont {Bulsara}},
  \bibinfo {author} {\bibfnamefont {W.~L.}\ \bibnamefont {Ditto}}, \bibinfo
  {author} {\bibfnamefont {S.}~\bibnamefont {Sinha}}, \bibinfo {author}
  {\bibfnamefont {K.}~\bibnamefont {Murali}}, \ and\ \bibinfo {author}
  {\bibfnamefont {P.}~\bibnamefont {Mohanty}},\ }\bibfield  {title} {\emph
  {\bibinfo {title} {A noise-assisted reprogrammable nanomechanical logic
  gate},\ }}\href@noop {} {\bibfield  {journal} {\bibinfo  {journal} {Nano
  Lett.}\ }\textbf {\bibinfo {volume} {10}},\ \bibinfo {pages} {1168} (\bibinfo
  {year} {2010})}\BibitemShut {NoStop}%
\bibitem [{\citenamefont {Tadokoro}\ and\ \citenamefont
  {Tanaka}(2021)}]{tadokoro2021highly}%
  \BibitemOpen
  \bibfield  {author} {\bibinfo {author} {\bibfnamefont {Y.}~\bibnamefont
  {Tadokoro}}\ and\ \bibinfo {author} {\bibfnamefont {H.}~\bibnamefont
  {Tanaka}},\ }\bibfield  {title} {\emph {\bibinfo {title} {Highly sensitive
  implementation of logic gates with a nonlinear nanomechanical resonator},\
  }}\href@noop {} {\bibfield  {journal} {\bibinfo  {journal} {Physical Review
  Applied}\ }\textbf {\bibinfo {volume} {15}},\ \bibinfo {pages} {024058}
  (\bibinfo {year} {2021})}\BibitemShut {NoStop}%
\bibitem [{\citenamefont {Rugar}\ and\ \citenamefont
  {Gr{\"u}tter}(1991)}]{rugar1991mechanical}%
  \BibitemOpen
  \bibfield  {author} {\bibinfo {author} {\bibfnamefont {D.}~\bibnamefont
  {Rugar}}\ and\ \bibinfo {author} {\bibfnamefont {P.}~\bibnamefont
  {Gr{\"u}tter}},\ }\bibfield  {title} {\emph {\bibinfo {title} {Mechanical
  parametric amplification and thermomechanical noise squeezing},\ }}\href@noop
  {} {\bibfield  {journal} {\bibinfo  {journal} {Phys. Rev. Lett.}\ }\textbf
  {\bibinfo {volume} {67}},\ \bibinfo {pages} {699} (\bibinfo {year}
  {1991})}\BibitemShut {NoStop}%
\bibitem [{\citenamefont {Karabalin}\ \emph {et~al.}(2011)\citenamefont
  {Karabalin}, \citenamefont {Lifshitz}, \citenamefont {Cross}, \citenamefont
  {Matheny}, \citenamefont {Masmanidis},\ and\ \citenamefont
  {Roukes}}]{karabalin2011signal}%
  \BibitemOpen
  \bibfield  {author} {\bibinfo {author} {\bibfnamefont {R.}~\bibnamefont
  {Karabalin}}, \bibinfo {author} {\bibfnamefont {R.}~\bibnamefont {Lifshitz}},
  \bibinfo {author} {\bibfnamefont {M.}~\bibnamefont {Cross}}, \bibinfo
  {author} {\bibfnamefont {M.}~\bibnamefont {Matheny}}, \bibinfo {author}
  {\bibfnamefont {S.}~\bibnamefont {Masmanidis}}, \ and\ \bibinfo {author}
  {\bibfnamefont {M.}~\bibnamefont {Roukes}},\ }\bibfield  {title} {\emph
  {\bibinfo {title} {Signal amplification by sensitive control of bifurcation
  topology},\ }}\href@noop {} {\bibfield  {journal} {\bibinfo  {journal} {Phys.
  Rev. Lett.}\ }\textbf {\bibinfo {volume} {106}},\ \bibinfo {pages} {094102}
  (\bibinfo {year} {2011})}\BibitemShut {NoStop}%
\bibitem [{\citenamefont {Papariello}\ \emph {et~al.}(2016)\citenamefont
  {Papariello}, \citenamefont {Zilberberg}, \citenamefont {Eichler},\ and\
  \citenamefont {Chitra}}]{papariello2016ultrasensitive}%
  \BibitemOpen
  \bibfield  {author} {\bibinfo {author} {\bibfnamefont {L.}~\bibnamefont
  {Papariello}}, \bibinfo {author} {\bibfnamefont {O.}~\bibnamefont
  {Zilberberg}}, \bibinfo {author} {\bibfnamefont {A.}~\bibnamefont {Eichler}},
  \ and\ \bibinfo {author} {\bibfnamefont {R.}~\bibnamefont {Chitra}},\
  }\bibfield  {title} {\emph {\bibinfo {title} {Ultrasensitive hysteretic force
  sensing with parametric nonlinear oscillators},\ }}\href@noop {} {\bibfield
  {journal} {\bibinfo  {journal} {Phys. Rev. E}\ }\textbf {\bibinfo {volume}
  {94}},\ \bibinfo {pages} {022201} (\bibinfo {year} {2016})}\BibitemShut
  {NoStop}%
\bibitem [{\citenamefont {Chowdhury}\ \emph {et~al.}(2020)\citenamefont
  {Chowdhury}, \citenamefont {Clerc}, \citenamefont {Barbay}, \citenamefont
  {Robert-Philip},\ and\ \citenamefont {Braive}}]{chowdhury2020weak}%
  \BibitemOpen
  \bibfield  {author} {\bibinfo {author} {\bibfnamefont {A.}~\bibnamefont
  {Chowdhury}}, \bibinfo {author} {\bibfnamefont {M.~G.}\ \bibnamefont
  {Clerc}}, \bibinfo {author} {\bibfnamefont {S.}~\bibnamefont {Barbay}},
  \bibinfo {author} {\bibfnamefont {I.}~\bibnamefont {Robert-Philip}}, \ and\
  \bibinfo {author} {\bibfnamefont {R.}~\bibnamefont {Braive}},\ }\bibfield
  {title} {\emph {\bibinfo {title} {Weak signal enhancement by nonlinear
  resonance control in a forced nano-electromechanical resonator},\
  }}\href@noop {} {\bibfield  {journal} {\bibinfo  {journal} {Nat. Commun.}\
  }\textbf {\bibinfo {volume} {11}},\ \bibinfo {pages} {1} (\bibinfo {year}
  {2020})}\BibitemShut {NoStop}%
\bibitem [{\citenamefont {Huber}\ \emph {et~al.}(2020)\citenamefont {Huber},
  \citenamefont {Rastelli}, \citenamefont {Seitner}, \citenamefont {K{\"o}lbl},
  \citenamefont {Belzig}, \citenamefont {Dykman},\ and\ \citenamefont
  {Weig}}]{huber2020spectral}%
  \BibitemOpen
  \bibfield  {author} {\bibinfo {author} {\bibfnamefont {J.~S.}\ \bibnamefont
  {Huber}}, \bibinfo {author} {\bibfnamefont {G.}~\bibnamefont {Rastelli}},
  \bibinfo {author} {\bibfnamefont {M.~J.}\ \bibnamefont {Seitner}}, \bibinfo
  {author} {\bibfnamefont {J.}~\bibnamefont {K{\"o}lbl}}, \bibinfo {author}
  {\bibfnamefont {W.}~\bibnamefont {Belzig}}, \bibinfo {author} {\bibfnamefont
  {M.~I.}\ \bibnamefont {Dykman}}, \ and\ \bibinfo {author} {\bibfnamefont
  {E.~M.}\ \bibnamefont {Weig}},\ }\bibfield  {title} {\emph {\bibinfo {title}
  {Spectral evidence of squeezing of a weakly damped driven nanomechanical
  mode},\ }}\href@noop {} {\bibfield  {journal} {\bibinfo  {journal} {Phys.
  Rev. X}\ }\textbf {\bibinfo {volume} {10}},\ \bibinfo {pages} {021066}
  (\bibinfo {year} {2020})}\BibitemShut {NoStop}%
\bibitem [{\citenamefont {Camerer}\ \emph {et~al.}(2011)\citenamefont
  {Camerer}, \citenamefont {Korppi}, \citenamefont {J{\"o}ckel}, \citenamefont
  {Hunger}, \citenamefont {H{\"a}nsch},\ and\ \citenamefont
  {Treutlein}}]{camerer2011realization}%
  \BibitemOpen
  \bibfield  {author} {\bibinfo {author} {\bibfnamefont {S.}~\bibnamefont
  {Camerer}}, \bibinfo {author} {\bibfnamefont {M.}~\bibnamefont {Korppi}},
  \bibinfo {author} {\bibfnamefont {A.}~\bibnamefont {J{\"o}ckel}}, \bibinfo
  {author} {\bibfnamefont {D.}~\bibnamefont {Hunger}}, \bibinfo {author}
  {\bibfnamefont {T.~W.}\ \bibnamefont {H{\"a}nsch}}, \ and\ \bibinfo {author}
  {\bibfnamefont {P.}~\bibnamefont {Treutlein}},\ }\bibfield  {title} {\emph
  {\bibinfo {title} {Realization of an optomechanical interface between
  ultracold atoms and a membrane},\ }}\href@noop {} {\bibfield  {journal}
  {\bibinfo  {journal} {Phys. Rev. Lett.}\ }\textbf {\bibinfo {volume} {107}},\
  \bibinfo {pages} {223001} (\bibinfo {year} {2011})}\BibitemShut {NoStop}%
\bibitem [{\citenamefont {Purdy}\ \emph {et~al.}(2013)\citenamefont {Purdy},
  \citenamefont {Peterson},\ and\ \citenamefont
  {Regal}}]{purdy2013observation}%
  \BibitemOpen
  \bibfield  {author} {\bibinfo {author} {\bibfnamefont {T.~P.}\ \bibnamefont
  {Purdy}}, \bibinfo {author} {\bibfnamefont {R.~W.}\ \bibnamefont {Peterson}},
  \ and\ \bibinfo {author} {\bibfnamefont {C.}~\bibnamefont {Regal}},\
  }\bibfield  {title} {\emph {\bibinfo {title} {Observation of radiation
  pressure shot noise on a macroscopic object},\ }}\href@noop {} {\bibfield
  {journal} {\bibinfo  {journal} {Science}\ }\textbf {\bibinfo {volume}
  {339}},\ \bibinfo {pages} {801} (\bibinfo {year} {2013})}\BibitemShut
  {NoStop}%
\bibitem [{\citenamefont {Andrews}\ \emph {et~al.}(2014)\citenamefont
  {Andrews}, \citenamefont {Peterson}, \citenamefont {Purdy}, \citenamefont
  {Cicak}, \citenamefont {Simmonds}, \citenamefont {Regal},\ and\ \citenamefont
  {Lehnert}}]{andrews2014bidirectional}%
  \BibitemOpen
  \bibfield  {author} {\bibinfo {author} {\bibfnamefont {R.~W.}\ \bibnamefont
  {Andrews}}, \bibinfo {author} {\bibfnamefont {R.~W.}\ \bibnamefont
  {Peterson}}, \bibinfo {author} {\bibfnamefont {T.~P.}\ \bibnamefont {Purdy}},
  \bibinfo {author} {\bibfnamefont {K.}~\bibnamefont {Cicak}}, \bibinfo
  {author} {\bibfnamefont {R.~W.}\ \bibnamefont {Simmonds}}, \bibinfo {author}
  {\bibfnamefont {C.~A.}\ \bibnamefont {Regal}}, \ and\ \bibinfo {author}
  {\bibfnamefont {K.~W.}\ \bibnamefont {Lehnert}},\ }\bibfield  {title} {\emph
  {\bibinfo {title} {Bidirectional and efficient conversion between microwave
  and optical light},\ }}\href@noop {} {\bibfield  {journal} {\bibinfo
  {journal} {Nat. Phys.}\ }\textbf {\bibinfo {volume} {10}},\ \bibinfo {pages}
  {321} (\bibinfo {year} {2014})}\BibitemShut {NoStop}%
\bibitem [{\citenamefont {J{\"o}ckel}\ \emph {et~al.}(2015)\citenamefont
  {J{\"o}ckel}, \citenamefont {Faber}, \citenamefont {Kampschulte},
  \citenamefont {Korppi}, \citenamefont {Rakher},\ and\ \citenamefont
  {Treutlein}}]{jockel2015sympathetic}%
  \BibitemOpen
  \bibfield  {author} {\bibinfo {author} {\bibfnamefont {A.}~\bibnamefont
  {J{\"o}ckel}}, \bibinfo {author} {\bibfnamefont {A.}~\bibnamefont {Faber}},
  \bibinfo {author} {\bibfnamefont {T.}~\bibnamefont {Kampschulte}}, \bibinfo
  {author} {\bibfnamefont {M.}~\bibnamefont {Korppi}}, \bibinfo {author}
  {\bibfnamefont {M.~T.}\ \bibnamefont {Rakher}}, \ and\ \bibinfo {author}
  {\bibfnamefont {P.}~\bibnamefont {Treutlein}},\ }\bibfield  {title} {\emph
  {\bibinfo {title} {Sympathetic cooling of a membrane oscillator in a hybrid
  mechanical--atomic system},\ }}\href@noop {} {\bibfield  {journal} {\bibinfo
  {journal} {Nat. Nanotechnol.}\ }\textbf {\bibinfo {volume} {10}},\ \bibinfo
  {pages} {55} (\bibinfo {year} {2015})}\BibitemShut {NoStop}%
\bibitem [{\citenamefont {Xu}\ \emph {et~al.}(2016)\citenamefont {Xu},
  \citenamefont {Mason}, \citenamefont {Jiang},\ and\ \citenamefont
  {Harris}}]{xu2016topological}%
  \BibitemOpen
  \bibfield  {author} {\bibinfo {author} {\bibfnamefont {H.}~\bibnamefont
  {Xu}}, \bibinfo {author} {\bibfnamefont {D.}~\bibnamefont {Mason}}, \bibinfo
  {author} {\bibfnamefont {L.}~\bibnamefont {Jiang}}, \ and\ \bibinfo {author}
  {\bibfnamefont {J.}~\bibnamefont {Harris}},\ }\bibfield  {title} {\emph
  {\bibinfo {title} {Topological energy transfer in an optomechanical system
  with exceptional points},\ }}\href@noop {} {\bibfield  {journal} {\bibinfo
  {journal} {Nature}\ }\textbf {\bibinfo {volume} {537}},\ \bibinfo {pages}
  {80} (\bibinfo {year} {2016})}\BibitemShut {NoStop}%
\bibitem [{\citenamefont {Karg}\ \emph {et~al.}(2020)\citenamefont {Karg},
  \citenamefont {Gouraud}, \citenamefont {Ngai}, \citenamefont {Schmid},
  \citenamefont {Hammerer},\ and\ \citenamefont {Treutlein}}]{karg2020light}%
  \BibitemOpen
  \bibfield  {author} {\bibinfo {author} {\bibfnamefont {T.~M.}\ \bibnamefont
  {Karg}}, \bibinfo {author} {\bibfnamefont {B.}~\bibnamefont {Gouraud}},
  \bibinfo {author} {\bibfnamefont {C.~T.}\ \bibnamefont {Ngai}}, \bibinfo
  {author} {\bibfnamefont {G.-L.}\ \bibnamefont {Schmid}}, \bibinfo {author}
  {\bibfnamefont {K.}~\bibnamefont {Hammerer}}, \ and\ \bibinfo {author}
  {\bibfnamefont {P.}~\bibnamefont {Treutlein}},\ }\bibfield  {title} {\emph
  {\bibinfo {title} {Light-mediated strong coupling between a mechanical
  oscillator and atomic spins 1 meter apart},\ }}\href@noop {} {\bibfield
  {journal} {\bibinfo  {journal} {Science}\ }\textbf {\bibinfo {volume}
  {369}},\ \bibinfo {pages} {174} (\bibinfo {year} {2020})}\BibitemShut
  {NoStop}%
\bibitem [{\citenamefont {Nayfeh}\ and\ \citenamefont
  {Mook}(2008)}]{nayfeh2008nonlinear}%
  \BibitemOpen
  \bibfield  {author} {\bibinfo {author} {\bibfnamefont {A.~H.}\ \bibnamefont
  {Nayfeh}}\ and\ \bibinfo {author} {\bibfnamefont {D.~T.}\ \bibnamefont
  {Mook}},\ }\href@noop {} {\emph {\bibinfo {title} {Nonlinear oscillations}}}\
  (\bibinfo  {publisher} {John Wiley \& Sons},\ \bibinfo {year}
  {2008})\BibitemShut {NoStop}%
\bibitem [{\citenamefont {Lifshitz}\ and\ \citenamefont
  {Cross}(2008)}]{schuster2009reviews}%
  \BibitemOpen
  \bibfield  {author} {\bibinfo {author} {\bibfnamefont {R.}~\bibnamefont
  {Lifshitz}}\ and\ \bibinfo {author} {\bibfnamefont {M.~C.}\ \bibnamefont
  {Cross}},\ }\href@noop {} {\emph {\bibinfo {title} {Reviews of nonlinear
  dynamics and complexity}}},\ edited by\ \bibinfo {editor} {\bibfnamefont
  {H.~G.}\ \bibnamefont {Schuster}}\ (\bibinfo  {publisher} {John Wiley \&
  Sons},\ \bibinfo {year} {2008})\BibitemShut {NoStop}%
\bibitem [{\citenamefont {Westra}\ \emph {et~al.}(2010)\citenamefont {Westra},
  \citenamefont {Poot}, \citenamefont {Van Der~Zant},\ and\ \citenamefont
  {Venstra}}]{westra2010nonlinear}%
  \BibitemOpen
  \bibfield  {author} {\bibinfo {author} {\bibfnamefont {H.}~\bibnamefont
  {Westra}}, \bibinfo {author} {\bibfnamefont {M.}~\bibnamefont {Poot}},
  \bibinfo {author} {\bibfnamefont {H.}~\bibnamefont {Van Der~Zant}}, \ and\
  \bibinfo {author} {\bibfnamefont {W.}~\bibnamefont {Venstra}},\ }\bibfield
  {title} {\emph {\bibinfo {title} {Nonlinear modal interactions in
  clamped-clamped mechanical resonators},\ }}\href@noop {} {\bibfield
  {journal} {\bibinfo  {journal} {Phys. Rev. Lett.}\ }\textbf {\bibinfo
  {volume} {105}},\ \bibinfo {pages} {117205} (\bibinfo {year}
  {2010})}\BibitemShut {NoStop}%
\bibitem [{\citenamefont {Manevich}\ and\ \citenamefont
  {Manevitch}(2005)}]{manevich2005mechanics}%
  \BibitemOpen
  \bibfield  {author} {\bibinfo {author} {\bibfnamefont {A.~I.}\ \bibnamefont
  {Manevich}}\ and\ \bibinfo {author} {\bibfnamefont {L.~I.}\ \bibnamefont
  {Manevitch}},\ }\href@noop {} {\emph {\bibinfo {title} {The mechanics of
  nonlinear systems with internal resonances}}}\ (\bibinfo  {publisher} {World
  Scientific},\ \bibinfo {year} {2005})\BibitemShut {NoStop}%
\bibitem [{\citenamefont {Mangussi}\ and\ \citenamefont
  {Zanette}(2016)}]{mangussi2016internal}%
  \BibitemOpen
  \bibfield  {author} {\bibinfo {author} {\bibfnamefont {F.}~\bibnamefont
  {Mangussi}}\ and\ \bibinfo {author} {\bibfnamefont {D.~H.}\ \bibnamefont
  {Zanette}},\ }\bibfield  {title} {\emph {\bibinfo {title} {Internal resonance
  in a vibrating beam: a zoo of nonlinear resonance peaks},\ }}\href@noop {}
  {\bibfield  {journal} {\bibinfo  {journal} {PloS one}\ }\textbf {\bibinfo
  {volume} {11}},\ \bibinfo {pages} {e0162365} (\bibinfo {year}
  {2016})}\BibitemShut {NoStop}%
\bibitem [{\citenamefont {Shoshani}\ \emph {et~al.}(2017)\citenamefont
  {Shoshani}, \citenamefont {Shaw},\ and\ \citenamefont
  {Dykman}}]{shoshani2017anomalous}%
  \BibitemOpen
  \bibfield  {author} {\bibinfo {author} {\bibfnamefont {O.}~\bibnamefont
  {Shoshani}}, \bibinfo {author} {\bibfnamefont {S.}~\bibnamefont {Shaw}}, \
  and\ \bibinfo {author} {\bibfnamefont {M.}~\bibnamefont {Dykman}},\
  }\bibfield  {title} {\emph {\bibinfo {title} {Anomalous decay of
  nanomechanical modes going through nonlinear resonance},\ }}\href@noop {}
  {\bibfield  {journal} {\bibinfo  {journal} {Sci. Rep.}\ }\textbf {\bibinfo
  {volume} {7}},\ \bibinfo {pages} {18091} (\bibinfo {year}
  {2017})}\BibitemShut {NoStop}%
\bibitem [{\citenamefont {G{\"u}ttinger}\ \emph {et~al.}(2017)\citenamefont
  {G{\"u}ttinger}, \citenamefont {Noury}, \citenamefont {Weber}, \citenamefont
  {Eriksson}, \citenamefont {Lagoin}, \citenamefont {Moser}, \citenamefont
  {Eichler}, \citenamefont {Wallraff}, \citenamefont {Isacsson},\ and\
  \citenamefont {Bachtold}}]{guttinger2017energy}%
  \BibitemOpen
  \bibfield  {author} {\bibinfo {author} {\bibfnamefont {J.}~\bibnamefont
  {G{\"u}ttinger}}, \bibinfo {author} {\bibfnamefont {A.}~\bibnamefont
  {Noury}}, \bibinfo {author} {\bibfnamefont {P.}~\bibnamefont {Weber}},
  \bibinfo {author} {\bibfnamefont {A.~M.}\ \bibnamefont {Eriksson}}, \bibinfo
  {author} {\bibfnamefont {C.}~\bibnamefont {Lagoin}}, \bibinfo {author}
  {\bibfnamefont {J.}~\bibnamefont {Moser}}, \bibinfo {author} {\bibfnamefont
  {C.}~\bibnamefont {Eichler}}, \bibinfo {author} {\bibfnamefont
  {A.}~\bibnamefont {Wallraff}}, \bibinfo {author} {\bibfnamefont
  {A.}~\bibnamefont {Isacsson}}, \ and\ \bibinfo {author} {\bibfnamefont
  {A.}~\bibnamefont {Bachtold}},\ }\bibfield  {title} {\emph {\bibinfo {title}
  {Energy-dependent path of dissipation in nanomechanical resonators},\
  }}\href@noop {} {\bibfield  {journal} {\bibinfo  {journal} {Nat.
  Nanotechnol.}\ } (\bibinfo {year} {2017})}\BibitemShut {NoStop}%
\bibitem [{\citenamefont {Chen}\ \emph {et~al.}(2017)\citenamefont {Chen},
  \citenamefont {Zanette}, \citenamefont {Czaplewski}, \citenamefont {Shaw},\
  and\ \citenamefont {L{\'o}pez}}]{chen2017direct}%
  \BibitemOpen
  \bibfield  {author} {\bibinfo {author} {\bibfnamefont {C.}~\bibnamefont
  {Chen}}, \bibinfo {author} {\bibfnamefont {D.~H.}\ \bibnamefont {Zanette}},
  \bibinfo {author} {\bibfnamefont {D.~A.}\ \bibnamefont {Czaplewski}},
  \bibinfo {author} {\bibfnamefont {S.}~\bibnamefont {Shaw}}, \ and\ \bibinfo
  {author} {\bibfnamefont {D.}~\bibnamefont {L{\'o}pez}},\ }\bibfield  {title}
  {\emph {\bibinfo {title} {Direct observation of coherent energy transfer in
  nonlinear micromechanical oscillators},\ }}\href@noop {} {\bibfield
  {journal} {\bibinfo  {journal} {Nat. Commun.}\ }\textbf {\bibinfo {volume}
  {8}},\ \bibinfo {pages} {15523} (\bibinfo {year} {2017})}\BibitemShut
  {NoStop}%
\bibitem [{\citenamefont {Yang}\ \emph {et~al.}(2019)\citenamefont {Yang},
  \citenamefont {Rochau}, \citenamefont {Huber}, \citenamefont {Brieussel},
  \citenamefont {Rastelli}, \citenamefont {Weig},\ and\ \citenamefont
  {Scheer}}]{yang2019spatial}%
  \BibitemOpen
  \bibfield  {author} {\bibinfo {author} {\bibfnamefont {F.}~\bibnamefont
  {Yang}}, \bibinfo {author} {\bibfnamefont {F.}~\bibnamefont {Rochau}},
  \bibinfo {author} {\bibfnamefont {J.~S.}\ \bibnamefont {Huber}}, \bibinfo
  {author} {\bibfnamefont {A.}~\bibnamefont {Brieussel}}, \bibinfo {author}
  {\bibfnamefont {G.}~\bibnamefont {Rastelli}}, \bibinfo {author}
  {\bibfnamefont {E.~M.}\ \bibnamefont {Weig}}, \ and\ \bibinfo {author}
  {\bibfnamefont {E.}~\bibnamefont {Scheer}},\ }\bibfield  {title} {\emph
  {\bibinfo {title} {Spatial modulation of nonlinear flexural vibrations of
  membrane resonators},\ }}\href {\doibase 10.1103/PhysRevLett.122.154301}
  {\bibfield  {journal} {\bibinfo  {journal} {Phys. Rev. Lett.}\ }\textbf
  {\bibinfo {volume} {122}},\ \bibinfo {pages} {154301} (\bibinfo {year}
  {2019})}\BibitemShut {NoStop}%
\bibitem [{\citenamefont {Waitz}\ \emph {et~al.}(2012)\citenamefont {Waitz},
  \citenamefont {N{\"o}{\ss}ner}, \citenamefont {Hertkorn}, \citenamefont
  {Schecker},\ and\ \citenamefont {Scheer}}]{waitz2012mode}%
  \BibitemOpen
  \bibfield  {author} {\bibinfo {author} {\bibfnamefont {R.}~\bibnamefont
  {Waitz}}, \bibinfo {author} {\bibfnamefont {S.}~\bibnamefont
  {N{\"o}{\ss}ner}}, \bibinfo {author} {\bibfnamefont {M.}~\bibnamefont
  {Hertkorn}}, \bibinfo {author} {\bibfnamefont {O.}~\bibnamefont {Schecker}},
  \ and\ \bibinfo {author} {\bibfnamefont {E.}~\bibnamefont {Scheer}},\
  }\bibfield  {title} {\emph {\bibinfo {title} {Mode shape and dispersion
  relation of bending waves in thin silicon membranes},\ }}\href@noop {}
  {\bibfield  {journal} {\bibinfo  {journal} {Phys. Rev. B}\ }\textbf {\bibinfo
  {volume} {85}},\ \bibinfo {pages} {035324} (\bibinfo {year}
  {2012})}\BibitemShut {NoStop}%
\bibitem [{\citenamefont {Yang}\ \emph {et~al.}(2017)\citenamefont {Yang},
  \citenamefont {Waitz},\ and\ \citenamefont {Scheer}}]{yang2017quantitative}%
  \BibitemOpen
  \bibfield  {author} {\bibinfo {author} {\bibfnamefont {F.}~\bibnamefont
  {Yang}}, \bibinfo {author} {\bibfnamefont {R.}~\bibnamefont {Waitz}}, \ and\
  \bibinfo {author} {\bibfnamefont {E.}~\bibnamefont {Scheer}},\ }\href@noop {}
  {\bibinfo {title} {Quantitative determination of the mechanical properties of
  nanomembrane resonators by vibrometry in continuous light},\ } (\bibinfo
  {year} {2017}),\ \Eprint {http://arxiv.org/abs/1704.05328} {arXiv:1704.05328
  [cond-mat.mes-hall]} \BibitemShut {NoStop}%
\bibitem [{\citenamefont {Petitgrand}\ \emph {et~al.}(2001)\citenamefont
  {Petitgrand}, \citenamefont {Yahiaoui}, \citenamefont {Danaie}, \citenamefont
  {Bosseboeuf},\ and\ \citenamefont {Gilles}}]{petitgrand20013d}%
  \BibitemOpen
  \bibfield  {author} {\bibinfo {author} {\bibfnamefont {S.}~\bibnamefont
  {Petitgrand}}, \bibinfo {author} {\bibfnamefont {R.}~\bibnamefont
  {Yahiaoui}}, \bibinfo {author} {\bibfnamefont {K.}~\bibnamefont {Danaie}},
  \bibinfo {author} {\bibfnamefont {A.}~\bibnamefont {Bosseboeuf}}, \ and\
  \bibinfo {author} {\bibfnamefont {J.}~\bibnamefont {Gilles}},\ }\bibfield
  {title} {\emph {\bibinfo {title} {3d measurement of micromechanical devices
  vibration mode shapes with a stroboscopic interferometric microscope},\
  }}\href@noop {} {\bibfield  {journal} {\bibinfo  {journal} {Opt. Lasers
  Eng.}\ }\textbf {\bibinfo {volume} {36}},\ \bibinfo {pages} {77} (\bibinfo
  {year} {2001})}\BibitemShut {NoStop}%
\bibitem [{SM()}]{SM}%
  \BibitemOpen
  \href@noop {} {}\bibinfo {note} {Supplemental Material for details of the
  sample preparation and the measurement methods, characterization measurements
  of the membrane under study, examples of sub-harmonic parametric resonance
  phenomena observed for different flexural modes, the identifications of mode
  frequency mixing, the energy decay results of ringdown process and the
  detailed description of the theoretical models.}\BibitemShut {Stop}%
\bibitem [{Note1()}]{Note1}%
  \BibitemOpen
  \bibinfo {note} {Note that the inclusion of higher-order terms causes still
  three possible solutions in the parameter regime of the experiment, two
  stable and an unstable one, as in the Duffing case, not shown in in Fig. \ref
  {fig:persistent_curve}(b).}\BibitemShut {Stop}%
\bibitem [{Note2()}]{Note2}%
  \BibitemOpen
  \bibinfo {note} {The first overtone at $2\protect \tmspace +\thinmuskip
  {.1667em}\omega _d$ is also present and couples to another higher-order mode,
  see Supplemental Material \cite {SM} for further examples.}\BibitemShut
  {Stop}%
\bibitem [{\citenamefont {Antoni}\ \emph {et~al.}(2013)\citenamefont {Antoni},
  \citenamefont {Makles}, \citenamefont {Braive}, \citenamefont {Briant},
  \citenamefont {Cohadon}, \citenamefont {Sagnes}, \citenamefont
  {Robert-Philip},\ and\ \citenamefont {Heidmann}}]{antoni2013nonlinear}%
  \BibitemOpen
  \bibfield  {author} {\bibinfo {author} {\bibfnamefont {T.}~\bibnamefont
  {Antoni}}, \bibinfo {author} {\bibfnamefont {K.}~\bibnamefont {Makles}},
  \bibinfo {author} {\bibfnamefont {R.}~\bibnamefont {Braive}}, \bibinfo
  {author} {\bibfnamefont {T.}~\bibnamefont {Briant}}, \bibinfo {author}
  {\bibfnamefont {P.-F.}\ \bibnamefont {Cohadon}}, \bibinfo {author}
  {\bibfnamefont {I.}~\bibnamefont {Sagnes}}, \bibinfo {author} {\bibfnamefont
  {I.}~\bibnamefont {Robert-Philip}}, \ and\ \bibinfo {author} {\bibfnamefont
  {A.}~\bibnamefont {Heidmann}},\ }\bibfield  {title} {\emph {\bibinfo {title}
  {Nonlinear mechanics with suspended nanomembranes},\ }}\href@noop {}
  {\bibfield  {journal} {\bibinfo  {journal} {EPL (Europhysics Letters)}\
  }\textbf {\bibinfo {volume} {100}},\ \bibinfo {pages} {68005} (\bibinfo
  {year} {2013})}\BibitemShut {NoStop}%
\bibitem [{\citenamefont {Polunin}\ \emph {et~al.}(2016)\citenamefont
  {Polunin}, \citenamefont {Yang}, \citenamefont {Dykman}, \citenamefont
  {Kenny},\ and\ \citenamefont {Shaw}}]{polunin2016characterization}%
  \BibitemOpen
  \bibfield  {author} {\bibinfo {author} {\bibfnamefont {P.~M.}\ \bibnamefont
  {Polunin}}, \bibinfo {author} {\bibfnamefont {Y.}~\bibnamefont {Yang}},
  \bibinfo {author} {\bibfnamefont {M.~I.}\ \bibnamefont {Dykman}}, \bibinfo
  {author} {\bibfnamefont {T.~W.}\ \bibnamefont {Kenny}}, \ and\ \bibinfo
  {author} {\bibfnamefont {S.~W.}\ \bibnamefont {Shaw}},\ }\bibfield  {title}
  {\emph {\bibinfo {title} {Characterization of mems resonator nonlinearities
  using the ringdown response},\ }}\href@noop {} {\bibfield  {journal}
  {\bibinfo  {journal} {J. Microelectromech. Syst.}\ }\textbf {\bibinfo
  {volume} {25}},\ \bibinfo {pages} {297} (\bibinfo {year} {2016})}\BibitemShut
  {NoStop}%
\end{thebibliography}%


\begin{thebibliography}{11}%
\makeatletter
\providecommand \@ifxundefined [1]{%
 \@ifx{#1\undefined}
}%
\providecommand \@ifnum [1]{%
 \ifnum #1\expandafter \@firstoftwo
 \else \expandafter \@secondoftwo
 \fi
}%
\providecommand \@ifx [1]{%
 \ifx #1\expandafter \@firstoftwo
 \else \expandafter \@secondoftwo
 \fi
}%
\providecommand \natexlab [1]{#1}%
\providecommand \enquote  [1]{``#1''}%
\providecommand \bibnamefont  [1]{#1}%
\providecommand \bibfnamefont [1]{#1}%
\providecommand \citenamefont [1]{#1}%
\providecommand \href@noop [0]{\@secondoftwo}%
\providecommand \href [0]{\begingroup \@sanitize@url \@href}%
\providecommand \@href[1]{\@@startlink{#1}\@@href}%
\providecommand \@@href[1]{\endgroup#1\@@endlink}%
\providecommand \@sanitize@url [0]{\catcode `\\12\catcode `\$12\catcode
  `\&12\catcode `\#12\catcode `\^12\catcode `\_12\catcode `\%12\relax}%
\providecommand \@@startlink[1]{}%
\providecommand \@@endlink[0]{}%
\providecommand \url  [0]{\begingroup\@sanitize@url \@url }%
\providecommand \@url [1]{\endgroup\@href {#1}{\urlprefix }}%
\providecommand \urlprefix  [0]{URL }%
\providecommand \Eprint [0]{\href }%
\providecommand \doibase [0]{http://dx.doi.org/}%
\providecommand \selectlanguage [0]{\@gobble}%
\providecommand \bibinfo  [0]{\@secondoftwo}%
\providecommand \bibfield  [0]{\@secondoftwo}%
\providecommand \translation [1]{[#1]}%
\providecommand \BibitemOpen [0]{}%
\providecommand \bibitemStop [0]{}%
\providecommand \bibitemNoStop [0]{.\EOS\space}%
\providecommand \EOS [0]{\spacefactor3000\relax}%
\providecommand \BibitemShut  [1]{\csname bibitem#1\endcsname}%
\let\auto@bib@innerbib\@empty
\bibitem [{\citenamefont {Yang}\ \emph {et~al.}(2019)\citenamefont {Yang},
  \citenamefont {Rochau}, \citenamefont {Huber}, \citenamefont {Brieussel},
  \citenamefont {Rastelli}, \citenamefont {Weig},\ and\ \citenamefont
  {Scheer}}]{yang2019spatial}%
  \BibitemOpen
  \bibfield  {author} {\bibinfo {author} {\bibfnamefont {F.}~\bibnamefont
  {Yang}}, \bibinfo {author} {\bibfnamefont {F.}~\bibnamefont {Rochau}},
  \bibinfo {author} {\bibfnamefont {J.~S.}\ \bibnamefont {Huber}}, \bibinfo
  {author} {\bibfnamefont {A.}~\bibnamefont {Brieussel}}, \bibinfo {author}
  {\bibfnamefont {G.}~\bibnamefont {Rastelli}}, \bibinfo {author}
  {\bibfnamefont {E.~M.}\ \bibnamefont {Weig}}, \ and\ \bibinfo {author}
  {\bibfnamefont {E.}~\bibnamefont {Scheer}},\ }\bibfield  {title} {\emph
  {\bibinfo {title} {Spatial modulation of nonlinear flexural vibrations of
  membrane resonators},\ }}\href {\doibase 10.1103/PhysRevLett.122.154301}
  {\bibfield  {journal} {\bibinfo  {journal} {Phys. Rev. Lett.}\ }\textbf
  {\bibinfo {volume} {122}},\ \bibinfo {pages} {154301} (\bibinfo {year}
  {2019})}\BibitemShut {NoStop}%
\bibitem [{\citenamefont {Yang}\ \emph {et~al.}(2017)\citenamefont {Yang},
  \citenamefont {Waitz},\ and\ \citenamefont {Scheer}}]{yang2017quantitative}%
  \BibitemOpen
  \bibfield  {author} {\bibinfo {author} {\bibfnamefont {F.}~\bibnamefont
  {Yang}}, \bibinfo {author} {\bibfnamefont {R.}~\bibnamefont {Waitz}}, \ and\
  \bibinfo {author} {\bibfnamefont {E.}~\bibnamefont {Scheer}},\ }\href@noop {}
  {\bibinfo {title} {Quantitative determination of the mechanical properties of
  nanomembrane resonators by vibrometry in continuous light},\ } (\bibinfo
  {year} {2017}),\ \Eprint {http://arxiv.org/abs/1704.05328} {arXiv:1704.05328
  [cond-mat.mes-hall]} \BibitemShut {NoStop}%
\bibitem [{\citenamefont {Waitz}\ \emph {et~al.}(2012)\citenamefont {Waitz},
  \citenamefont {N{\"o}{\ss}ner}, \citenamefont {Hertkorn}, \citenamefont
  {Schecker},\ and\ \citenamefont {Scheer}}]{waitz2012mode}%
  \BibitemOpen
  \bibfield  {author} {\bibinfo {author} {\bibfnamefont {R.}~\bibnamefont
  {Waitz}}, \bibinfo {author} {\bibfnamefont {S.}~\bibnamefont
  {N{\"o}{\ss}ner}}, \bibinfo {author} {\bibfnamefont {M.}~\bibnamefont
  {Hertkorn}}, \bibinfo {author} {\bibfnamefont {O.}~\bibnamefont {Schecker}},
  \ and\ \bibinfo {author} {\bibfnamefont {E.}~\bibnamefont {Scheer}},\
  }\bibfield  {title} {\emph {\bibinfo {title} {Mode shape and dispersion
  relation of bending waves in thin silicon membranes},\ }}\href@noop {}
  {\bibfield  {journal} {\bibinfo  {journal} {Phys. Rev. B}\ }\textbf {\bibinfo
  {volume} {85}},\ \bibinfo {pages} {035324} (\bibinfo {year}
  {2012})}\BibitemShut {NoStop}%
\bibitem [{\citenamefont {Landau}\ and\ \citenamefont
  {Lifshitz}(2013)}]{landau2013course}%
  \BibitemOpen
  \bibfield  {author} {\bibinfo {author} {\bibfnamefont {L.~D.}\ \bibnamefont
  {Landau}}\ and\ \bibinfo {author} {\bibfnamefont {E.~M.}\ \bibnamefont
  {Lifshitz}},\ }\href@noop {} {\emph {\bibinfo {title} {Course of theoretical
  physics}}}\ (\bibinfo  {publisher} {Elsevier},\ \bibinfo {year}
  {2013})\BibitemShut {NoStop}%
\bibitem [{\citenamefont {Zhang}\ \emph {et~al.}(2015)\citenamefont {Zhang},
  \citenamefont {Waitz}, \citenamefont {Yang}, \citenamefont {Lutz},
  \citenamefont {Angelova}, \citenamefont {G{\"o}lzh{\"a}user},\ and\
  \citenamefont {Scheer}}]{zhang2015vibrational}%
  \BibitemOpen
  \bibfield  {author} {\bibinfo {author} {\bibfnamefont {X.}~\bibnamefont
  {Zhang}}, \bibinfo {author} {\bibfnamefont {R.}~\bibnamefont {Waitz}},
  \bibinfo {author} {\bibfnamefont {F.}~\bibnamefont {Yang}}, \bibinfo {author}
  {\bibfnamefont {C.}~\bibnamefont {Lutz}}, \bibinfo {author} {\bibfnamefont
  {P.}~\bibnamefont {Angelova}}, \bibinfo {author} {\bibfnamefont
  {A.}~\bibnamefont {G{\"o}lzh{\"a}user}}, \ and\ \bibinfo {author}
  {\bibfnamefont {E.}~\bibnamefont {Scheer}},\ }\bibfield  {title} {\emph
  {\bibinfo {title} {Vibrational modes of ultrathin carbon nanomembrane
  mechanical resonators},\ }}\href@noop {} {\bibfield  {journal} {\bibinfo
  {journal} {Appl. Phys. Lett.}\ }\textbf {\bibinfo {volume} {106}},\ \bibinfo
  {pages} {063107} (\bibinfo {year} {2015})}\BibitemShut {NoStop}%
\bibitem [{\citenamefont {Waitz}\ \emph {et~al.}(2015)\citenamefont {Waitz},
  \citenamefont {Lutz}, \citenamefont {N{\"o}{\ss}ner}, \citenamefont
  {Hertkorn},\ and\ \citenamefont {Scheer}}]{waitz2015spatially}%
  \BibitemOpen
  \bibfield  {author} {\bibinfo {author} {\bibfnamefont {R.}~\bibnamefont
  {Waitz}}, \bibinfo {author} {\bibfnamefont {C.}~\bibnamefont {Lutz}},
  \bibinfo {author} {\bibfnamefont {S.}~\bibnamefont {N{\"o}{\ss}ner}},
  \bibinfo {author} {\bibfnamefont {M.}~\bibnamefont {Hertkorn}}, \ and\
  \bibinfo {author} {\bibfnamefont {E.}~\bibnamefont {Scheer}},\ }\bibfield
  {title} {\emph {\bibinfo {title} {Spatially resolved measurement of the
  stress tensor in thin membranes using bending waves},\ }}\href@noop {}
  {\bibfield  {journal} {\bibinfo  {journal} {Phys. Rev. Applied}\ }\textbf
  {\bibinfo {volume} {3}},\ \bibinfo {pages} {044002} (\bibinfo {year}
  {2015})}\BibitemShut {NoStop}%
\bibitem [{\citenamefont {Antoni}\ \emph {et~al.}(2013)\citenamefont {Antoni},
  \citenamefont {Makles}, \citenamefont {Braive}, \citenamefont {Briant},
  \citenamefont {Cohadon}, \citenamefont {Sagnes}, \citenamefont
  {Robert-Philip},\ and\ \citenamefont {Heidmann}}]{antoni2013nonlinear}%
  \BibitemOpen
  \bibfield  {author} {\bibinfo {author} {\bibfnamefont {T.}~\bibnamefont
  {Antoni}}, \bibinfo {author} {\bibfnamefont {K.}~\bibnamefont {Makles}},
  \bibinfo {author} {\bibfnamefont {R.}~\bibnamefont {Braive}}, \bibinfo
  {author} {\bibfnamefont {T.}~\bibnamefont {Briant}}, \bibinfo {author}
  {\bibfnamefont {P.-F.}\ \bibnamefont {Cohadon}}, \bibinfo {author}
  {\bibfnamefont {I.}~\bibnamefont {Sagnes}}, \bibinfo {author} {\bibfnamefont
  {I.}~\bibnamefont {Robert-Philip}}, \ and\ \bibinfo {author} {\bibfnamefont
  {A.}~\bibnamefont {Heidmann}},\ }\bibfield  {title} {\emph {\bibinfo {title}
  {Nonlinear mechanics with suspended nanomembranes},\ }}\href@noop {}
  {\bibfield  {journal} {\bibinfo  {journal} {EPL (Europhysics Letters)}\
  }\textbf {\bibinfo {volume} {100}},\ \bibinfo {pages} {68005} (\bibinfo
  {year} {2013})}\BibitemShut {NoStop}%
\bibitem [{\citenamefont {Polunin}\ \emph {et~al.}(2016)\citenamefont
  {Polunin}, \citenamefont {Yang}, \citenamefont {Dykman}, \citenamefont
  {Kenny},\ and\ \citenamefont {Shaw}}]{polunin2016characterization}%
  \BibitemOpen
  \bibfield  {author} {\bibinfo {author} {\bibfnamefont {P.~M.}\ \bibnamefont
  {Polunin}}, \bibinfo {author} {\bibfnamefont {Y.}~\bibnamefont {Yang}},
  \bibinfo {author} {\bibfnamefont {M.~I.}\ \bibnamefont {Dykman}}, \bibinfo
  {author} {\bibfnamefont {T.~W.}\ \bibnamefont {Kenny}}, \ and\ \bibinfo
  {author} {\bibfnamefont {S.~W.}\ \bibnamefont {Shaw}},\ }\bibfield  {title}
  {\emph {\bibinfo {title} {Characterization of mems resonator nonlinearities
  using the ringdown response},\ }}\href@noop {} {\bibfield  {journal}
  {\bibinfo  {journal} {J. Microelectromech. Syst.}\ }\textbf {\bibinfo
  {volume} {25}},\ \bibinfo {pages} {297} (\bibinfo {year} {2016})}\BibitemShut
  {NoStop}%
\bibitem [{\citenamefont {Lifshitz}\ and\ \citenamefont
  {Cross}(2008)}]{schuster2009reviews}%
  \BibitemOpen
  \bibfield  {author} {\bibinfo {author} {\bibfnamefont {R.}~\bibnamefont
  {Lifshitz}}\ and\ \bibinfo {author} {\bibfnamefont {M.~C.}\ \bibnamefont
  {Cross}},\ }\href@noop {} {\emph {\bibinfo {title} {Reviews of nonlinear
  dynamics and complexity}}},\ edited by\ \bibinfo {editor} {\bibfnamefont
  {H.~G.}\ \bibnamefont {Schuster}}\ (\bibinfo  {publisher} {John Wiley \&
  Sons},\ \bibinfo {year} {2008})\BibitemShut {NoStop}%
\bibitem [{\citenamefont {G{\"u}ttinger}\ \emph {et~al.}(2017)\citenamefont
  {G{\"u}ttinger}, \citenamefont {Noury}, \citenamefont {Weber}, \citenamefont
  {Eriksson}, \citenamefont {Lagoin}, \citenamefont {Moser}, \citenamefont
  {Eichler}, \citenamefont {Wallraff}, \citenamefont {Isacsson},\ and\
  \citenamefont {Bachtold}}]{guttinger2017energy}%
  \BibitemOpen
  \bibfield  {author} {\bibinfo {author} {\bibfnamefont {J.}~\bibnamefont
  {G{\"u}ttinger}}, \bibinfo {author} {\bibfnamefont {A.}~\bibnamefont
  {Noury}}, \bibinfo {author} {\bibfnamefont {P.}~\bibnamefont {Weber}},
  \bibinfo {author} {\bibfnamefont {A.~M.}\ \bibnamefont {Eriksson}}, \bibinfo
  {author} {\bibfnamefont {C.}~\bibnamefont {Lagoin}}, \bibinfo {author}
  {\bibfnamefont {J.}~\bibnamefont {Moser}}, \bibinfo {author} {\bibfnamefont
  {C.}~\bibnamefont {Eichler}}, \bibinfo {author} {\bibfnamefont
  {A.}~\bibnamefont {Wallraff}}, \bibinfo {author} {\bibfnamefont
  {A.}~\bibnamefont {Isacsson}}, \ and\ \bibinfo {author} {\bibfnamefont
  {A.}~\bibnamefont {Bachtold}},\ }\bibfield  {title} {\emph {\bibinfo {title}
  {Energy-dependent path of dissipation in nanomechanical resonators},\
  }}\href@noop {} {\bibfield  {journal} {\bibinfo  {journal} {Nat.
  Nanotechnol.}\ } (\bibinfo {year} {2017})}\BibitemShut {NoStop}%
\bibitem [{\citenamefont {Chen}\ \emph {et~al.}(2017)\citenamefont {Chen},
  \citenamefont {Zanette}, \citenamefont {Czaplewski}, \citenamefont {Shaw},\
  and\ \citenamefont {L{\'o}pez}}]{chen2017direct}%
  \BibitemOpen
  \bibfield  {author} {\bibinfo {author} {\bibfnamefont {C.}~\bibnamefont
  {Chen}}, \bibinfo {author} {\bibfnamefont {D.~H.}\ \bibnamefont {Zanette}},
  \bibinfo {author} {\bibfnamefont {D.~A.}\ \bibnamefont {Czaplewski}},
  \bibinfo {author} {\bibfnamefont {S.}~\bibnamefont {Shaw}}, \ and\ \bibinfo
  {author} {\bibfnamefont {D.}~\bibnamefont {L{\'o}pez}},\ }\bibfield  {title}
  {\emph {\bibinfo {title} {Direct observation of coherent energy transfer in
  nonlinear micromechanical oscillators},\ }}\href@noop {} {\bibfield
  {journal} {\bibinfo  {journal} {Nat. Commun.}\ }\textbf {\bibinfo {volume}
  {8}},\ \bibinfo {pages} {15523} (\bibinfo {year} {2017})}\BibitemShut
  {NoStop}%
\end{thebibliography}%

\end{document}


\title{Supplemental Material for ``Persistent response in ultra-strongly driven mechanical membrane resonators"}
%
\author{Fan Yang}
\affiliation{Fachbereich Physik, Universit{\"a}t Konstanz, 78457 Konstanz, Germany}
\author{Felicitas Hellbach}
\affiliation{Fachbereich Physik, Universit{\"a}t Konstanz, 78457 Konstanz, Germany}
 \author{Felix Rochau}
 \affiliation{Fachbereich Physik, Universit{\"a}t Konstanz, 78457 Konstanz, Germany}
 \author{Wolfgang Belzig}
 \affiliation{Fachbereich Physik, Universit{\"a}t Konstanz, 78457 Konstanz, Germany}
\author{Eva M. Weig}
\affiliation{Fachbereich Physik, Universit{\"a}t Konstanz, 78457 Konstanz, Germany}
\affiliation{Fakultät für Elektrotechnik und Informationstechnik, Technische Universit{\"a}t M{\"u}nchen, Germany}
\author{Gianluca Rastelli}
 \email{gianluca.rastelli@ino.cnr.it}
\affiliation{INO-CNR BEC Center and Dipartimento di Fisica, Universit{\`a} di Trento, 38123 Povo, Italy}
\affiliation{Fachbereich Physik, Universit{\"a}t Konstanz, 78457 Konstanz, Germany}
\author{Elke Scheer}%
 \email{elke.scheer@uni-konstanz.de}
\affiliation{Fachbereich Physik, Universit{\"a}t Konstanz, 78457 Konstanz, Germany}%

\maketitle
%
%
\section{\label{sec:level1} Fabrication and measurement principles}
\subsection{\label{sec:level2} Fabrication}
\begin{figure*}[htbp]
   \includegraphics[width=0.85\textwidth]{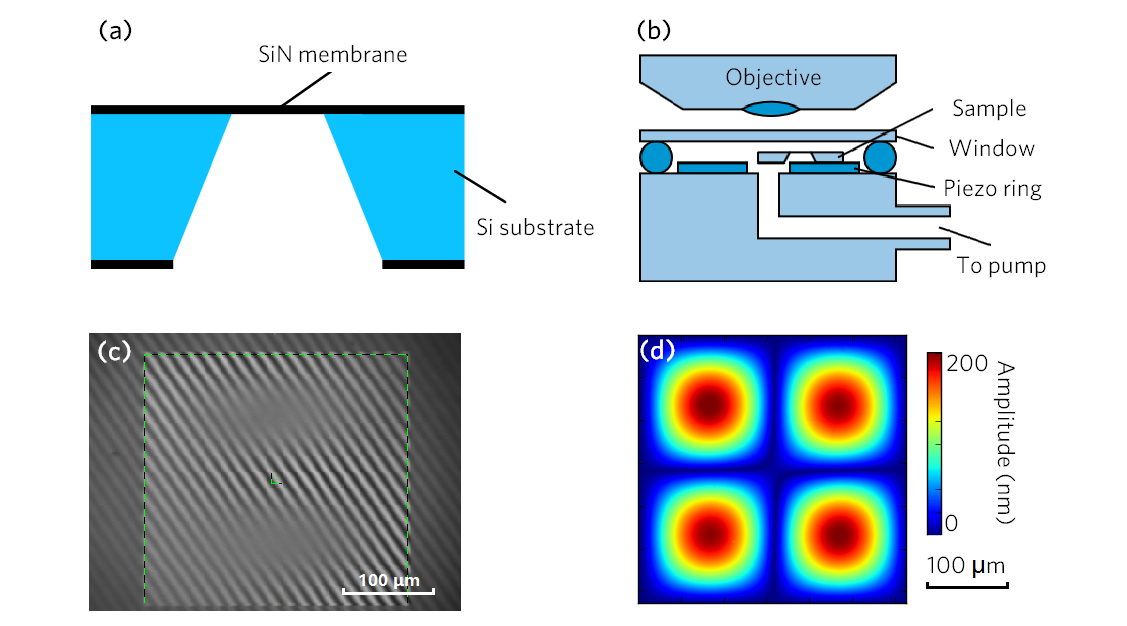}
   \caption[width=\textwidth]{SiN membrane sample and experimental setup of IWLI. (a) Cross section of a free-standing 500 nm thick SiN membrane supported by a 0.5 mm thick silicon substrate. (b) Sketch of the IWLI measurement setup showing the membrane chip, the piezo ring, the vacuum chamber and the objective. (c) Camera view of a 298 $\times$ 296 $\mu$m$^{2}$ membrane under the imaging interferometer objective using continuous light. The measurement area is indicated by the green frame. Here the membrane is in the (1,2) mode, as indicated by the blurring of the almost diagonal interference lines. (d) False color image of an amplitude profile of the membrane recorded at an excitation frequency of 909 kHz, corresponding to the (2,2) mode.}
   \label{fig:SM_mode_set}
 \end{figure*}

\noindent The silicon nitride (SiN) membranes are fabricated from a 0.5 mm thick commercial (100) silicon wafer, both sides of which are coated with $\mathrm{\sim}$500 nm thick LPCVD SiN. The membrane is fabricated on the front layer, and the backside layer serves as an etch mask. Laser ablation is used to open the etch mask with a typical size of 1 $\mathrm{\times}$ 1 mm$^{2}$. Using anisotropic etching in aqueous potassium nitride (KOH), a hole is etched through the openings of the mask. After about twenty hours the KOH solution reaches the topside layer and a membrane is formed, supported by a massive silicon frame. The cross section of a free-standing SiN membrane supported by the silicon substrate is presented in Fig. \ref{fig:SM_mode_set}(a). In this work, a 478 nm thick and 413.5 $\mathrm{\times}$ 393.5 $\mu $m$^{2}$ wide membrane is employed. The chip carrying the membrane is glued using 2-component adhesive with contact points (5 mm diameter) at each corner of the substrate to a piezo ring with 20 mm diameter and 5 mm thickness, as shown in Fig. \ref{fig:sample3D} \cite{yang2019spatial,yang2017quantitative}.
\subsection{\label{sec:level3} Imaging white light interferometry (IWLI)}
\begin{figure}[htbp]
   \includegraphics[width=0.5\linewidth]{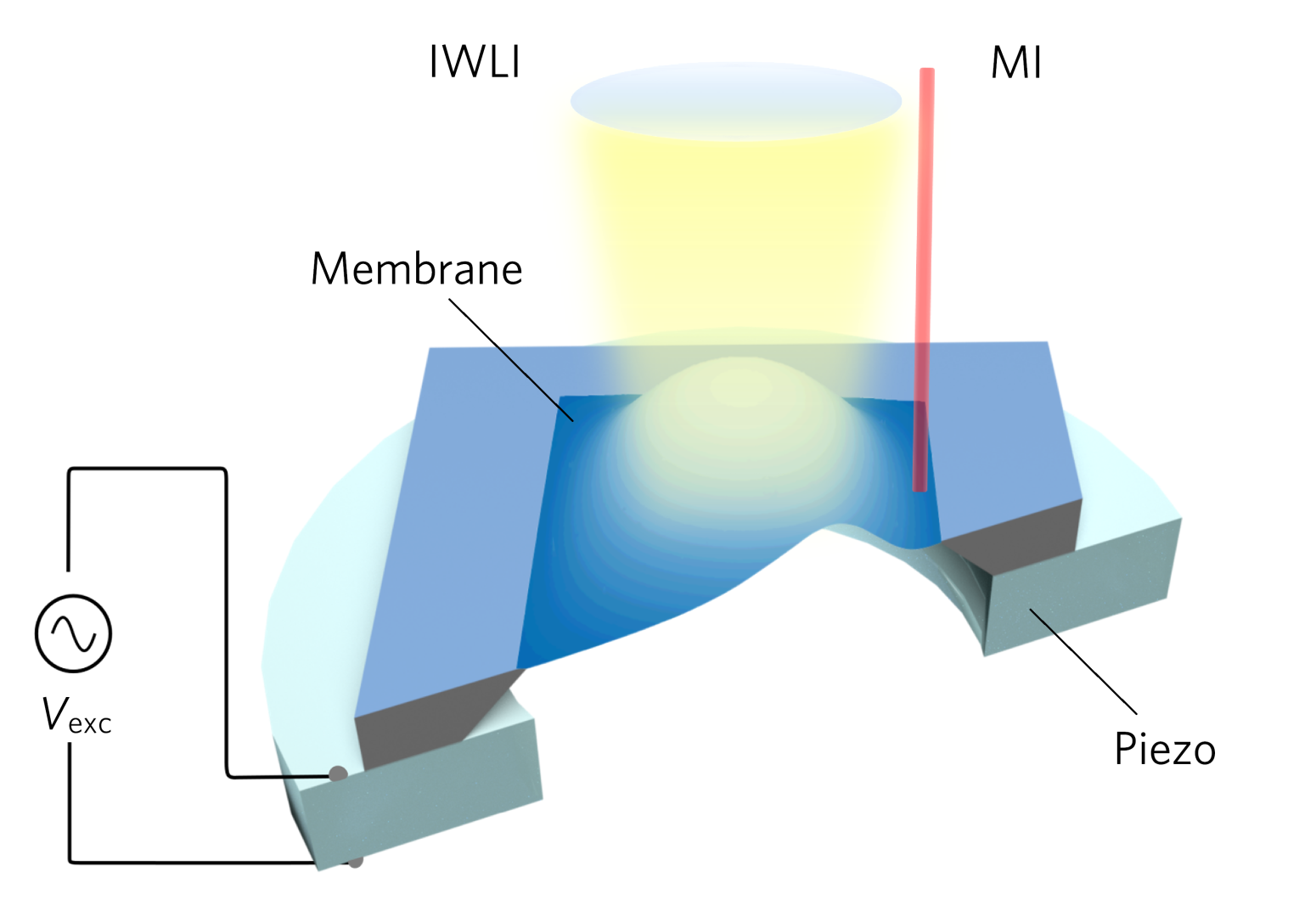}
   \caption[width=\textwidth]{Sketch of the experimental setups showing the membrane chip carrying a free-standing SiN membrane, the piezo ring and drive, the objective of the imaging white light interferometer (IWLI) and the laser beam of the Michelson interferometer (MI).}
   \label{fig:sample3D}
 \end{figure}

\noindent The sample is placed in a vacuum chamber connected to a pressure controller, see Fig. \ref{fig:SM_mode_set}(b), to ensure full control over the pressure of the surrounding atmosphere in a pressure range from \textit{p} = 0.001 mbar to atmospheric pressure. The measurements discussed in the present work have been performed at 2$\times 10^{-2}$ mbar. The surface of the membrane is observed by an imaging interferometer using different light sources, described in detail in Ref. \cite{waitz2012mode,yang2017quantitative}. The excitation voltage is applied using a sinusoidal function generator the phase of which can be locked to the stroboscopic light of the imaging white light interferometer, as shown in Fig. \ref{fig:sample3D}.
\noindent The observed interference pattern represents the surface profile of the sample as exemplified in Fig. \ref{fig:SM_mode_set}(c). The interference fringes can be used to quantitatively measure the vibrational amplitude. Figure \ref{fig:SM_mode_set}(d) shows a measurement example for the (2,2) mode. The vibrational deflection pattern can be measured by stroboscopic illumination and by continuous illumination. The deflection pattern measured by stroboscopic light contains vibrational phase information but is limited to the specific driving frequency with a locked phase. The continuous light can be applied without any phase lock-in, and thus provides deflection amplitude patterns. In this work, we utilized continuous light for recording the mode deflection patterns, stroboscopic light was utilized to measure the image shown in Fig. \ref{fig:SM_mode_set}(d).
\subsection{\label{sec:level4} Michelson interferometry}
\noindent A Michelson Interfermeter (MI) is utilized to measure the sweep-up and ring-down processes of the deflection at a certain point on the membrane area as shown in Fig. \ref{fig:sample3D}. A ($\Lambda$ = 1550 nm) laser beam is focused on a spot (\textit{d} $\approx$ 1 $\mu$m) on the free-standing membrane which is positioned using an \textit{xyz} piezo-positioning stage. The vibrations of the oscillating membrane modulate the reflected light, which is interfered with a phase reference (stabilized with a control bandwidth of 50 kHz). The detected power of the interference signal is proportional to the amplitude squared for deflections much smaller than the optical wavelength. For strong excitation voltages, this limit is hard to obey. At most positions the deflection amplitude is then much larger than a quarter of the laser wavelength used in the MI. To avoid ambiguities in the data interpretation we focus the laser on a point close to one edge of the membrane, where the amplitude remains well within one fringe of the MI signal. Frequency response spectra are measured using a fast lock-in amplifier with a bind width of 1~kHz. For the frequency-resolved ring-down measurements, the oscillations are recorded with an oscilloscope with a sampling rate of 10~MS/s. Then an fast Fourier transform (FFT) is performed on sets of 2000 sample points each. Integration around a particular oscillation frequency yields the separated energy decay traces shown in Figs. 3 and 4 of the main text. 
\section{\label{sec:level5} Characterization of mechanical properties}
\noindent By applying our customized VICL (Vibrometery In Continuous Light) \cite{landau2013course} and APSStro (Automated Phase-Shifting and Surface Measurement in Stroboscopic Light) measurement methods \cite{zhang2015vibrational, waitz2012mode} dispersion relations of the bending waves of the membrane are measured using IWLI. From this data, Young's modulus \textit{E} and the residual stress $\sigma$ are determined by fitting. The resonance frequencies as well as the mechanical $Q$ factors of different vibrational modes of this SiN membrane are quantitatively determined. For the (1,1) mode we find $\omega_{11}/2\pi~=~323.5~\textrm{kHz}$, $Q_{11} = 2\times 10^{4}$, $\textit{E}~=~240.2~\textrm{GPa}$, $\sigma~=~0.1261~\textrm{GPa}$. Note that temperature drifts induce frequency shifts in the order of 500 Hz/K. The temperature in the lab is stabilized with a precision of $\pm$ 1K. The IWLI and the MI are located in two different labs with slightly different average temperature. Thus, the absolute values of the eigenfrequencies may vary about $\pm$ 2.5 kHz from set-up to set-up. The absolute values indicated correspond to the ones measured with MI. The full width at half maximum (FWHM) of the (1,1) mode in the linear response regime is about 50 Hz. Table S1 gives an overview over the expected eigenfrequencies of the membrane, measured by MI and IWLI, and calculated using the formula:
\begin{align}
\label{eigen_f}
 \omega_{mn} \cong 2\pi\sqrt{{({\sigma}_{xx}{m^2}/{L^2_{\mathrm{w}}}+{\sigma}_{yy}{n^2}/{L^2_{\mathrm{h}}})}/{(4\rho )}}.   
\end{align}
Here the $\rho$ = 3.18 $\mathrm{\times}$ $10^{3}$ kg/m${}^{3}$ is the mass density, $L_{w}$ = 413.5 $\mu$m and $L_{h}$ = 393.5 $\mu$m are the edge lengths of the membrane, and \textit{m,n} denote the integer mode indices representing the number of antinodes. The observed temperature dependence of the eigenfrequencies is attributed to temperature-dependent stress tensor components $\sigma_{xx}$ and $\sigma_{yy}$ along the \textit{x} and \textit{y} axis, respectively. The values given in the table have been calculated using the values $\sigma_{xx}$ = 0.110 GPa and $\sigma_{yy}$ = 0.108 GPa determined by our spatially resolved measurement method described in Ref. \cite{waitz2015spatially}. Table \ref{tab:table01} lists the experimental and theoretical values of different flexural modes (\textit{m,n}) in kHz. The frequency values have been measured by IWLI and MI  (underlined).\\
As an example, we find $\omega_{11}/2\pi$ = 321 kHz and $\omega_{22}/2\pi$ = 646 kHz for the eigenfrequencies of (1,1) and (2,2) mode, respectively, both with quality factors in the order of 20000. Note that $\omega_{22}$ is not exactly twice the (1,1) mode frequency because of the non-zero bending rigidity and the slightly rectangular shape of the membrane. This feature will become important in the nonlinear regime.
\begin{table*}
  \caption{Experimental and theoretical values for the eigenfrequencies of different flexural modes. Measured eigenfrequencies of different flexural modes (\textit{m,n}) in kHz. The frequency values have been measured by MI (underlined) and IWLI. Calculated eigenfrequencies of different flexural modes (\textit{m},\textit{n}) in kHz. }
  \label{tab:table01} 
  \begin{ruledtabular}
  \begin{tabular}{ccccccccccc} \hline 
   & \multicolumn{5}{c}{Experimental} & \multicolumn{5}{c}{Theoretical} \\ \hline 
  Mode & 1 (\textit{n}) & 2 & 3 & 4 & 5 & 1 & 2 & 3 & 4 & 5 \\ \hline 
  1 (\textit{m}) & 321 & 500 & \underline{712} & 927 & 1174 & 325 & 507 & 714 & 930 & 1149 \\ \hline 
  2 & 517 & 646 & 818 & 1018 & 1233 & 520 & 649 & 821 & 1014 & 1218 \\ \hline 
  3 & \underline{747} & 837 & \underline{983} & 1148 & 1326 & 738 & 834 & 974 & 1141 & 1326 \\ \hline 
  4 & \underline{979} & \underline{1052} & 1160 & 1314 & 1495 & 963 & 1039 & 1154 & 1299 & 1463 \\ \hline 
  5 & 1188 & 1260 & 1351 & 1511 & 1640 & 1192 & 1254 & 1351 & 1477 & 1623 \\ \hline 
  \end{tabular}
  \end{ruledtabular}
\end{table*}
\section{\label{sec:level6} Persistent response behavior}
\begin{figure}[ht]
  \includegraphics[width=0.5\linewidth]{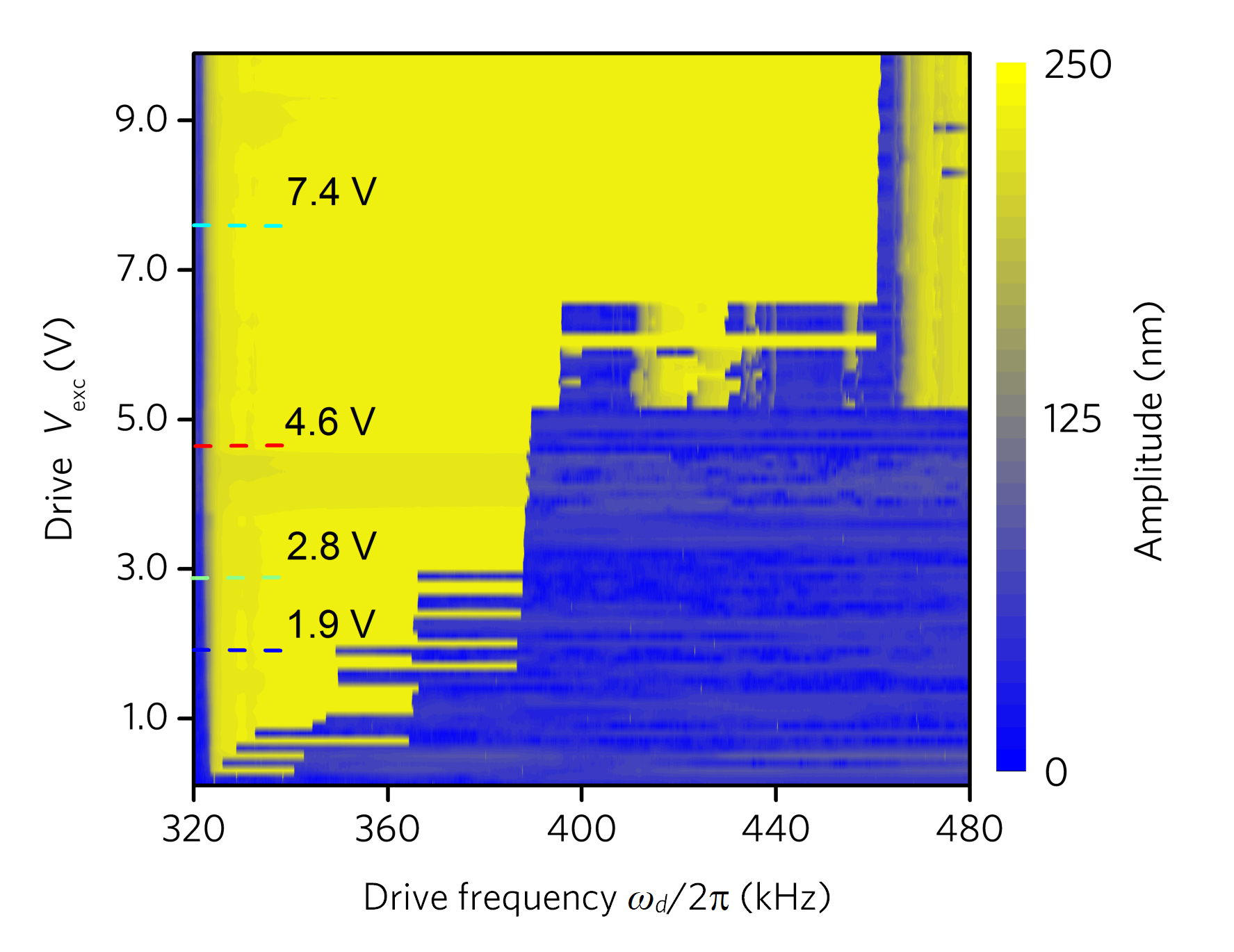}
  \caption[width=\linewidth]{Nonlinear response and amplitude saturation. False-color map of the nonlinear response of the membrane at frequency $\omega_{d}$ under increasing sinusoidal drive $V_\textrm{exc}$, measured by IWLI. The drive frequency is increased around the linear eigenfrequency $\omega_{11}$. The color bar shows the absolute amplitude response averaged over the entire membrane. The colored dashed lines mark the selected traces given in Fig. 1 of the main text.}
  \label{fig:2D_IWLI}
\end{figure}

\noindent In Fig. 1(a) of the main text we present four typical resonance curves to reveal the spatial modulation and flexural mode coupling behavior. In Fig. \ref{fig:2D_IWLI}, we present a complete 2D map of the amplitude to demonstrate the excitation-dependent response near the (1,1) mode. Like in Fig. 1(a) and Fig. 2(a) of the main text, the data is obtained by IWLI integrating over the whole membrane area. Figure \ref{fig:2D_IWLI} gives an overview of the behavior for 0.1 V $\mathrm{\le}$ $V_\textrm{exc}$ $\mathrm{\le}$ 9.5 V and in a frequency range of 320~kHz to 480 kHz as a false-color plot. After an initial fast increase of the amplitude, we observe a wide plateau that extends over a broad frequency range of up to 50\% of $\omega_{11}$. At the end of the plateau, the amplitude drops down abruptly by more than an order of magnitude. The dropping frequency shows a step-wise, non-monotonic, but reproducible dependence on $V_\textrm{exc}$, indicating the existence of at least two (metastable) vibrational states of the membrane. For excitation voltages 5.2 V $\mathrm{\le}$ $V_\textrm{exc}$ $\mathrm{\le}$ 6.4 V, additional nonlinear resonances are excited at around 410 kHz after the end of the persistent response plateau of the (1,1) mode. No other eigenfrequency beside the fundamental (1,1) mode is comprised in the plateau range. The driving force 5.2 V $\mathrm{\le}$ $V_\textrm{exc}$ $\mathrm{\le}$ 6.4 V is not strong enough to generate the persistent response plateau of the (1,1) mode between 400 kHz and 440 kHz, as seen by the fluctuating amplitude in this range. 

%
%

\newpage
\section{\label{sec:level7} Nonlinear response in the flexural mode coupling regime}

\subsection{Sub-harmonically driven nonlinear flexural modes}
\noindent Mode assignment of the ring-down measurements in Fig. 3 of the main text:
In Fig. 3(b), for sub-harmonically  driven (2,2) mode, we observe the following ring-down developments ($\rightarrow$: decays to): $\omega_d/2\pi$ = 322.0 kHz $\rightarrow$ $\omega_{11}/2\pi$ = 319.3 kHz, 2$\,\omega_d/2\pi$ = 644.0 kHz $\rightarrow$  $\omega_{22}/2\pi$ = 643.3 kHz and 3$\,\omega_d/2\pi$ = 966.0 kHz $\rightarrow$ $3\,\omega_{11}/2\pi$ = 957.9 kHz.
In Fig. 3(d), for sub-harmonically  driven (1,2) mode, we observe the following ring-down developments: $\omega_d/2\pi$ = 332.5 kHz $ \rightarrow $  $\omega_{11}/2\pi$ = 319.0 kHz, $\frac{3}{2}\,\omega_d/2\pi$ = 498.8 kHz $ \rightarrow $  $\omega_{12}/2\pi$ = 496.9 kHz, $\frac{6}{2}\,\omega_d/2\pi$ = 997.5 kHz $ \rightarrow $  2$\,\omega_{12}/2\pi$ = 993.8 kHz and $\frac{9}{2}\,\omega_d/2\pi$ = 1496.3 kHz $ \rightarrow $  3$\,\omega_{12}/2\pi$ = 1490.7 kHz. Higher frequency range are cut out and shown in Fig. \ref{fig:rd_cut_23} for the further discussion.\\

\begin{figure*}[htbp]
  \includegraphics[width=\linewidth]{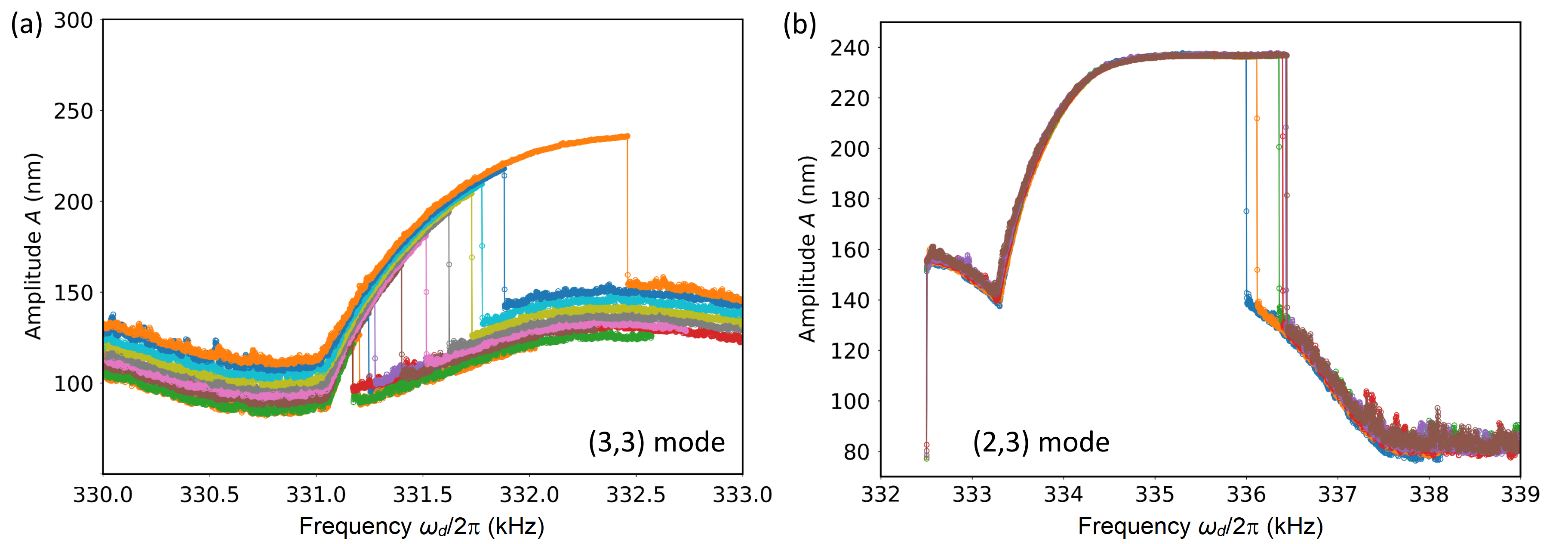}
  \caption[width=\linewidth]{Off-resonantly measured amplitude response. (a) Amplitude response of sub-harmonically driven (3,3) mode. Drives for each curves are: 4.62, 4.71, 4.76, 4.80, 4.85, 4.90, 4.94, 4.99, 5.17, 5.36, 5.54, 5.73 V, respectively. (b) Amplitude response of sub-harmonically driven (2,3) mode. Drives selected for each curves are 5.86, 5.91, 5.95, 6.00, 6.05, 6.09 V, respectively}
  \label{fig:res_33_23}
\end{figure*}

\noindent As an supplement to the (2,2) and (1,2) modes shown in the main text in Fig. 3, two additional examples, the (3,3) and (2,3) mode are shown here. We drive flexural modes sub-harmonically at frequencies higher than $\omega_{11}/2\pi$ to exclude the strong contribution of the (1,1) mode. \\

\noindent The curves in Fig. \ref{fig:res_33_23}(a) show the amplitude response of the sub-harmonically driven (3,3) mode under different excitations, the drive frequencies are around $\omega_d = \frac{1}{3}\,\omega_{33}/2\pi$ = 331 kHz. $V_\textrm{exc}$ is calibrated as: 4.62, 4.71, 4.76, 4.80, 4.85, 4.90, 4.94, 4.99, 5.17, 5.36, 5.54, 5.73 V, respectively. The ``Duffing-like" curves are attribute to the \textit{indirect parametric nonlinear interaction} ($m$ = $n$) character, the high background amplitude around 100 nm corresponds to the contribution of the (1,1) mode vibrating in low amplitude state.
The curves in Fig. \ref{fig:res_33_23}(b) show the amplitude response of the sub-harmonically driven (2,3) mode under different excitations again in the nonlinear regime, the drive frequencies are around $\omega_d/2\pi = \frac{2}{5}\,\omega_{23}/2\pi$ = 333.5 kHz. $V_\textrm{exc}$ is calibrated as 5.86, 5.91, 5.95, 6.00, 6.05, 6.09V, respectively. The curves shows a sub-harmonically driven \textit{direct parametric nonlinear interaction} ($m \neq n$) character. We discuss the parametric resonance curves of the (1,1) mode in the following.\\

\begin{figure*}[htbp]
 \centering
  \includegraphics[width=\linewidth]{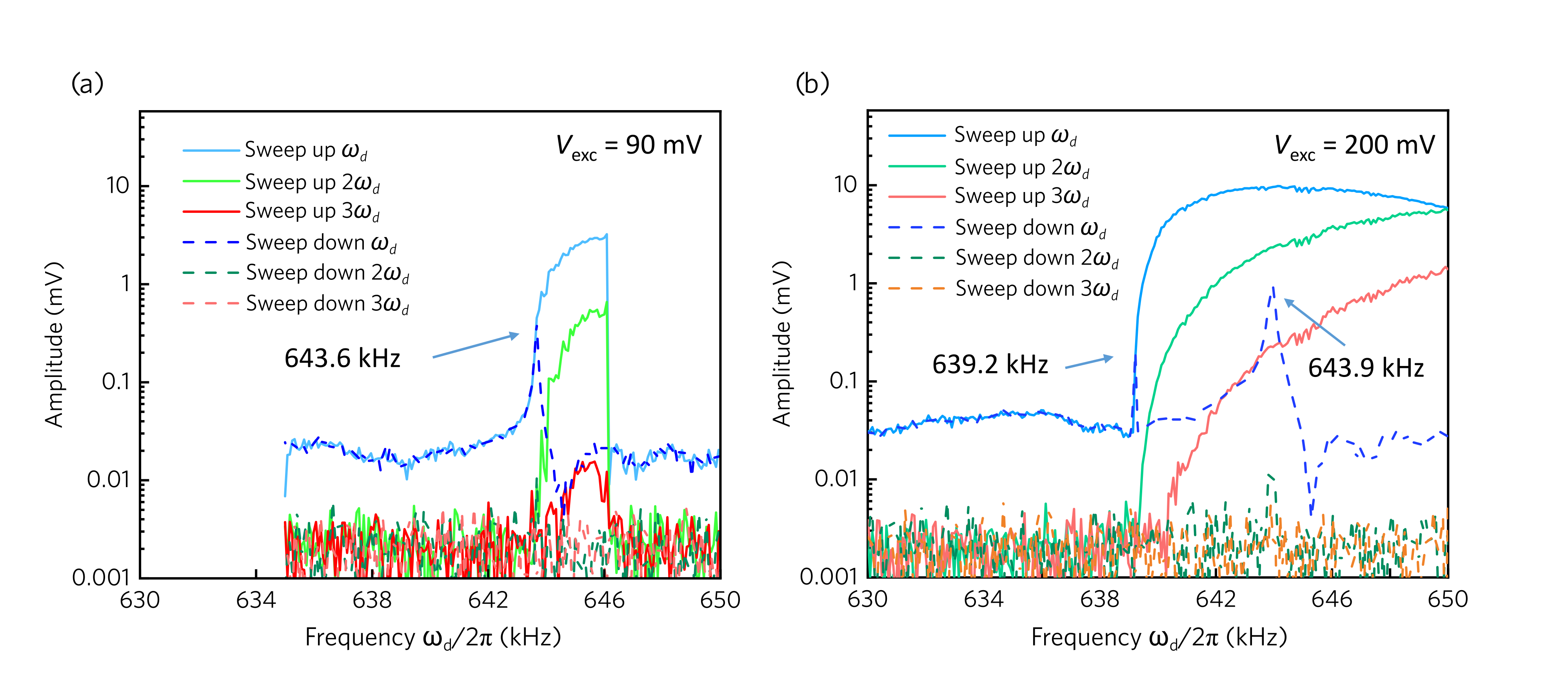}
  \caption{Parametrically driven resonance curve of the (1,1) mode (forward and backward frequency sweep) by using MI. The amplitude response of the $2\textsuperscript{nd}$ and $3\textsuperscript{rd}$ higher order components measured by lock-in are plotted as well. (a) The (2,2) mode is driven around $\omega_{22}/2\pi = 643.6~\textrm{kHz}$, with excitation $V_\textrm{exc}$ = 90 mV. (b) with $V_\textrm{exc}$ = 200 mV, the typical parametric resonance curve of (1,1) mode are recorded around 2$\,\omega_{11}/2\pi~=~639.2~\textrm{kHz}$. Two spikes corresponding to 2$\,\omega_{11}$ and $\omega_{22}$ can be found while sweeping backward at a distance of around 4.7 kHz.}
  \label{fig:parametric_11mode}
\end{figure*}

\begin{figure}[htbp]
  \includegraphics[width=0.5\linewidth]{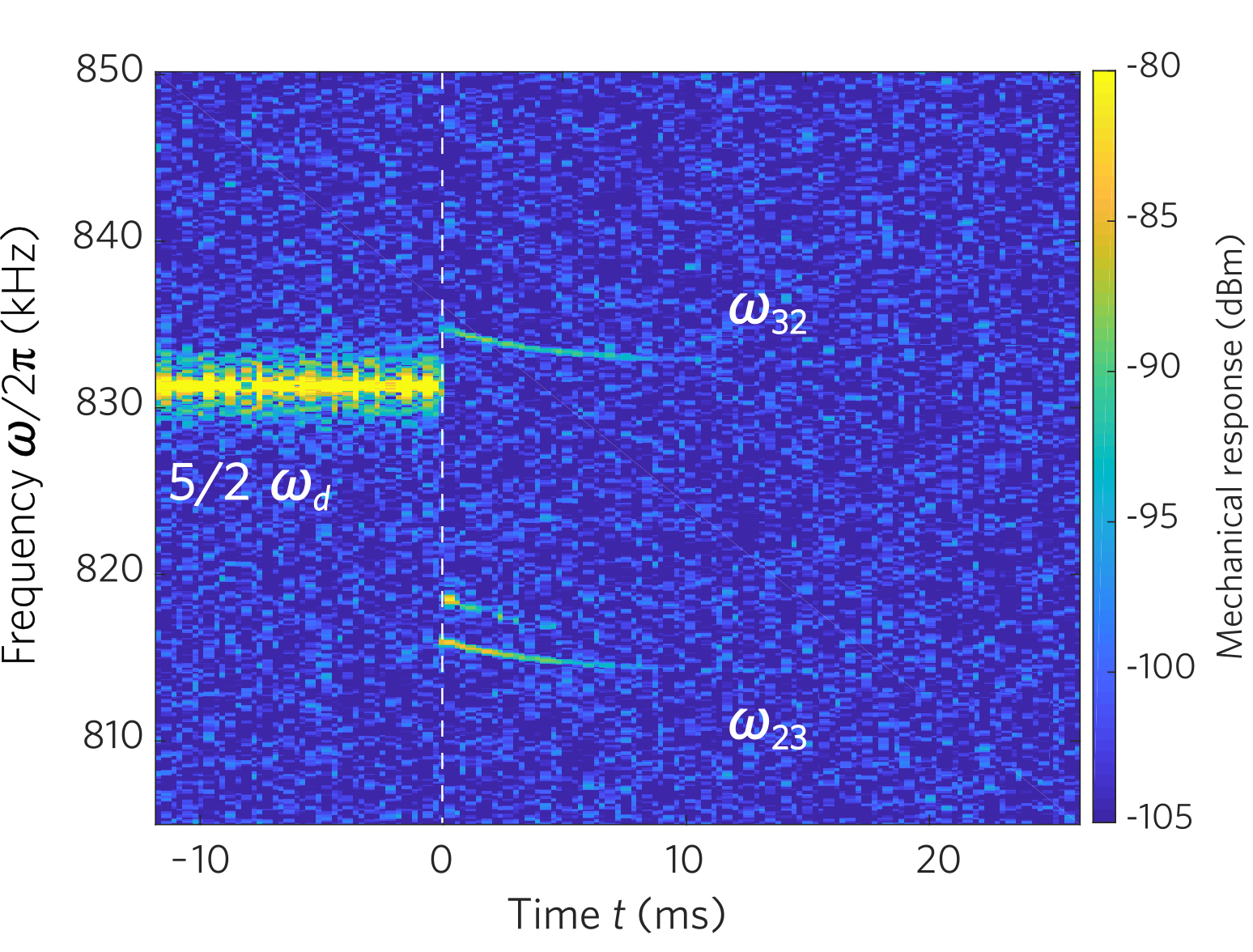}
  \caption{Ring-down frequency spectrum of sub-harmonically driven (1,2) mode for $V_\textrm{exc}$ = 7.0 V and $\omega_d/2\pi$ = 332.5 kHz, showing the frequency spectrum in the range of $\frac{5}{2}\,\omega_d$. The same measurement showing in Fig. 3(d) in the main text. The drive is switched off at $t~=~0~\textrm{ms}$. The mode frequency reading from the ring-down trace is $\omega_{23}/2\pi$ = 814 kHz and the $\omega_{32}/2\pi$ = 833 kHz.}
  \label{fig:rd_cut_23}
\end{figure}

\noindent For a quick overview of the parametric resonance, we perform a simple parametric excitation of (1,1) mode and measure the amplitude response by using MI as shown in Fig. \ref{fig:parametric_11mode}. The forward and backward frequency sweeps are recorded. The amplitude response of $2\textsuperscript{nd}$ and $3\textsuperscript{rd}$ higher order components measured by lock-in are plotted as well. The $\omega_d$ is swept around 2$\,\omega_{11}$ with excitation $V_\textrm{exc}$ = 90 mV and $V_\textrm{exc}$ = 200 mV. The corresponding resonance curves are plotted in (a) and (b). In Fig. \ref{fig:parametric_11mode} (a), only the (2,2) mode is excited at the $\omega_{22}/2\pi~=~643.6~\textrm{kHz}$, note here we didn't see any sign of parametric resonance of (1,1) mode. 
In the case of $V_\textrm{exc}$ = 200 mV, the $\omega_d$ is swept around 2$\,\omega_{11}$ again and the curves in panel (b) shows a strongly driven parametric resonance of (1,1) mode at 2$\,\omega_{11}/2\pi~=~639.2~\textrm{kHz}$ identified by the back-ward sweep. For the higher detuning, the (2,2) mode starts pumping up and coupling with the parametric driven (1,1) mode, we will not discuss this behavior into details here. 
The parametric resonance of (1,1) mode shown in panel (b) is similar to the resonance curves we measured by IWLI as shown in Fig. 3(d) in the main text for the (1,2) mode as well as in Fig. \ref{fig:res_33_23}(b) for the (2,3) mode. Differently, in the main text, we discussed the sub-harmonic parametrically driven response. The higher overtone of the driven nonlinear mode acts as a parametric driving source, transferring part of its energy, as long as the frequency of the overtone matches (close to) the twice the eigenfrequency of one specific flexural mode.\\

\noindent Similar to the sub-harmonic parametrically driven (1,2) mode shown in Fig. 3(d) in the main text, simultaneously, the $5\textsuperscript{th}$ overtone of the drive ($5\,\omega_d$) can parametrically excite the (2,3) and (3,2) mode as shown in Fig. \ref{fig:rd_cut_23} but with much weaker amplitude response. The ringdown power spectrum of the (2,3) and (3,2) mode shown in Fig. \ref{fig:rd_cut_23} is the same measurement shown in Fig. 3(d), cut out from frequency range of $\frac{5}{2}\,\omega_d$.\\ 

\noindent The fractional relation of sub-harmonic parametrically driven modes can be predicted as 2$\,\omega_{mn}$ = $(m + n) \cdot \omega_d$. The corresponding fractional frequency relations identified in the experiments and in theory are:\\

\noindent For the (1,2) mode:
\begin{itemize}
\centering
  \item $\omega_{12} \approx \frac{3}{2} \,\omega_d$,
  \item 2$\,\omega_{12} \approx \frac{6}{2} \,\omega_d$,
  \item 3$\,\omega_{12} \approx \frac{9}{2} \,\omega_d$,
\end{itemize}

\noindent for the (2,3) mode:
\begin{itemize}
\centering
  \item $\omega_{23} \approx \frac{5}{2} \,\omega_d$,
  \item $2\,\omega_{23} \approx \frac{10}{2} \,\omega_d$,
\end{itemize}

\noindent summarizing, for the $(m,n)$ mode:
\begin{itemize}
\centering
  \item $\omega_{mn} \approx \frac{m + n}{2}\,\omega_d$,
\end{itemize}

\noindent and for the $i\textsuperscript{th}$ overtone of the $(m,n)$ mode:
\begin{itemize}
\centering
  \item $i\,\omega_{mn} \approx \frac{i(m + n)}{2}\,\omega_d$.
\end{itemize}
\begin{figure}[t]
  \includegraphics[width=0.85\textwidth]{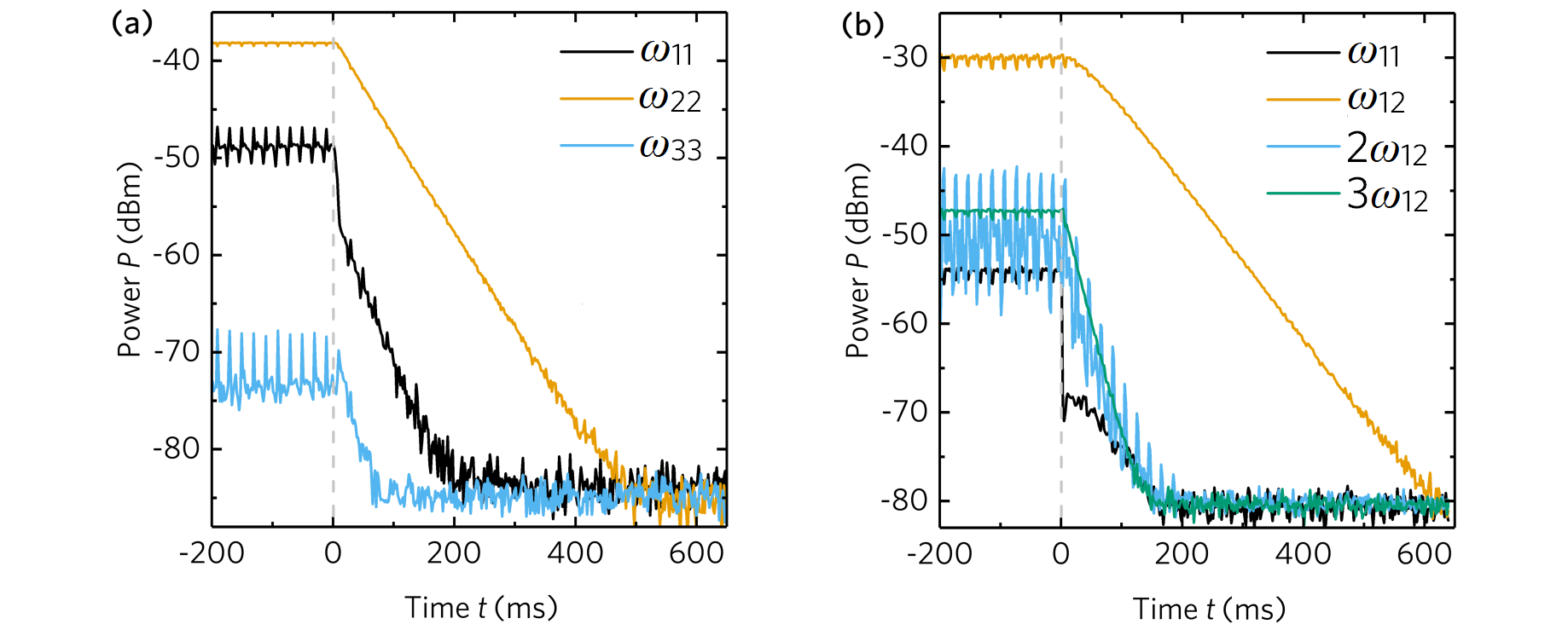}
  \caption[width=\textwidth]{Ring-down amplitude of sub-harmonically driven nonlinear flexural modes. Time resolved, and frequency integrated power spectra of the ringdown data shown in Fig. 3(b) and 3(d) of the main text for different sub-harmonically driven modes.(a) The ringdown trace of sub-harmonically driven (2,2), the ringdown traces plotted as $\omega_{11}$ black, $\omega_{22}$ yellow, $\omega_{33}$ blue). The trace of $3\,\omega_{11}$ is not shown due to the weak power. The ring-down starts at $t = 0$. (b) The ringdown trace of sub-harmonically driven (1,2), the ringdown traces plotted as $\omega_{11}$ black, $\omega_{12}$ yellow, 2$\,\omega_{12}$ blue, 3$\,\omega_{12}$ green). The ring-down starts at $t = 0$. The amplitude of the $\omega_{12}$ persists for around 30 ms and then decays exponentially, the 2$^\textrm{nd}$ and the 3$^\textrm{rd}$ overtone of sub-harmonically driven (1,2) mode decay immediately after switching off the drive and with a larger decay rate.}
  \label{fig:SM_sub_rd}
\end{figure}
\begin{figure}[htbp]
  \includegraphics[width=0.85\textwidth]{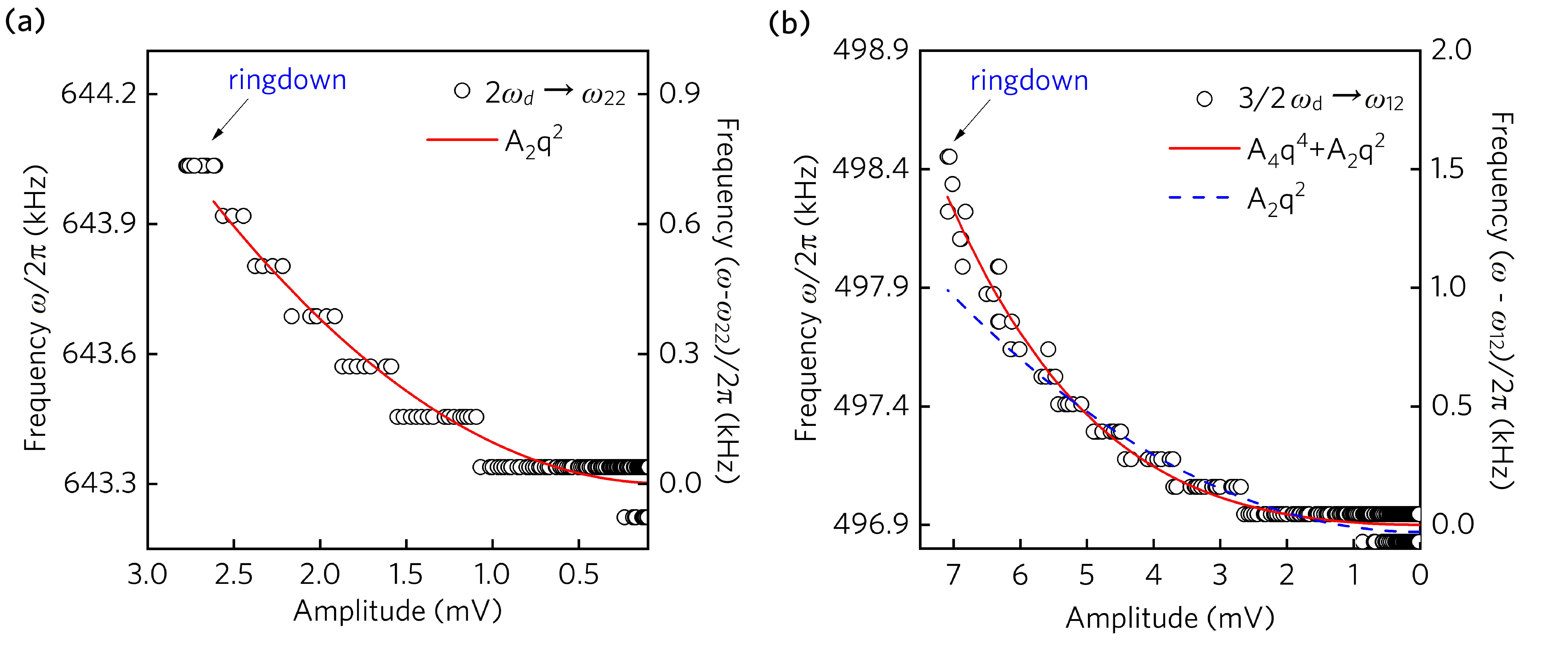}
  \caption[width=\textwidth]{(a) Time-resolved and frequency-integrated power vs. frequency decay for $2\,\omega_d \rightarrow \omega_{22}$ (black circles), fitted by a quadratic function (red line). Ring-down data shown in Fig. 3(b). (b) Same as (a) but for the data shown in Fig. 3(d) in the main text, i.e. $\frac{3}{2}\,\omega_d \rightarrow \omega_{12}$ (black circles), fitted by a quartic (red line) and a  quadratic (blue-dashed line) function.}
  \label{fig:SM_sub_dynamic}
\end{figure}
\noindent In the main text we show the driving nonlinear flexural modes sub-harmonically at a frequencies higher than $\omega_{11}$ to exclude the strong amplitude contribution of the (1,1) mode. Two examples, for the (2,2) and the (1,2) mode are showing in Fig. 3 (a,b) and (c,d) in the main text.
Figure~\ref{fig:SM_sub_rd} presents the time-resolved FFT power spectra obtained by integrating over a certain frequency range around the individual resonance frequencies.
For the sub-harmonically driven (2,2) mode, the power is mainly concentrated on the $\omega_{22}$, and characterized by the integrated ring-down power traces shown in Fig. \ref{fig:SM_sub_rd}(a), which shows a single-exponential decay of the (2,2) mode. The ring-down traces present anomalous decay behavior for $\omega_{11}$: A sudden drop of the amplitude of $\omega_{11}$ can be seen might due to the jumping of the driven state of $\omega_{11}$ to the back-bone trace of (1,1) mode in Duffing nonlinear regime when the power is off.\\ 

\noindent For the sub-harmonically driven (1,2) mode shown in the main text, the power is mainly concentrated on the $\omega_{12}$, and characterized by the integrated ring-down power traces shown in Fig. \ref{fig:SM_sub_rd}(b). The ring-down traces present anomalous decay for $\omega_{11}$, the sudden drop similar to the behavior in Fig. \ref{fig:SM_sub_rd} (a). Meanwhile, the anomalous decay for $\omega_{12}$ can be seen. its $2^\textrm{nd}$ overtone which reveals an oscillating faster decay. The signal powers of 2$\,\omega_{12}$ and 3$\,\omega_{12}$ are weaker than the one of $\omega_{12}$. The signal power of $\omega_{12}$ only starts decaying 30 ms after the drive has been switched off. Then starts a slow anomalous decay process between 30 ms to 70 ms. The decay traces of the $\omega_{12}$ power shows an exponential decay process 70 ms after the power has been switched off. The decay rates of the individual overtones are calculated by fitting an exponential function to the ring-down power integrated over the respective frequency range. The power oscillation of 2$\,\omega_{12}$ can be directly seen even at the steady state with power on. We argue that this extremely slow oscillation might be due to heat redistribution caused by thermoelastic processes. The response and the power oscillations of the 2$\,\omega_{12}$ mode are similar to the behavior of spatial modulation \cite{yang2019spatial}.
The  frequency-integrated ring-down amplitude vs. frequency decay traces of the experiment in the main text of Fig. 3(b) shown in Fig. \ref{fig:SM_sub_dynamic}(a) presents a typical quadratic decay of the (2,2) mode, indicating a cubic nonlinearity in the sub-harmonically driven (2,2) mode.
The frequency-integrated ring-down amplitude vs. frequency decay trace of the experiment in the main text of Fig. 3(d) is mainly concentrated on $\omega_{12}$, which can be fitted by a quartic decay relation (red line) shown in Fig. \ref{fig:SM_sub_dynamic}(b). The decay relation is indicating that the higher order nonlinearity exist in the system and might caused the large amplitude and the flatness of the amplitude response curves in the sub-harmonically driven 3/2 parametric resonance of (1,2) mode, see Fig. 3(c) in the main text. The best approximation of quadratic decay (blue-dashed line) is also shown.\\
%
%
%
%
%
%
\subsection{Flexural mode coupling and ring-down measurement}
\noindent We identified all the flexural modes and their overtones shown in Fig 4(a) in the main text. We observe the following main developments: $\omega_d/2\pi$ = 338.6 kHz $\rightarrow$  $\omega_{11}/2\pi$ = 320.2 kHz, 2$\,\omega_d/2\pi$ = 677.2 kHz $\rightarrow$  $\omega_{22}/2\pi$ = 645 kHz, 3$\,\omega_d/2\pi$ = 1015.8 kHz $\rightarrow$  $\omega_{33}/2\pi$ = 977.5 kHz, 4$\,\omega_d/2\pi$ = 1354.4 kHz $\rightarrow$  2$\,\omega_{22}/2\pi$\ =  1292 kHz and 2$\,\omega_{11}/2\pi$ + $\omega_{22}/2\pi$ = 1287 kHz, 6$\,\omega_d/2\pi$ = 2031.6 kHz $\rightarrow$  3$\,\omega_{22}/2\pi$ = 1946 kHz and 4$\,\omega_{12}/2\pi$  = 1939 kHz, 7$\,\omega_d/2\pi$ = 2370.2 kHz $\rightarrow$  2$\,\omega_{34}/2\pi$  = 2304 kHz (not plotted in (a)). The 5\textsuperscript{th} overtone splits up into the 2\textsuperscript{nd} overtone of the (2,3) and the (3,2) modes 5$\,\omega_d/2\pi$ = 1693 kHz $\rightarrow$  2$\,\omega_{23}/2\pi$ = 1636 kHz and 2$\,\omega_{32}/2\pi$ = 1624 kHz).\\

\begin{figure}[htbp]
  \includegraphics[width=\linewidth]{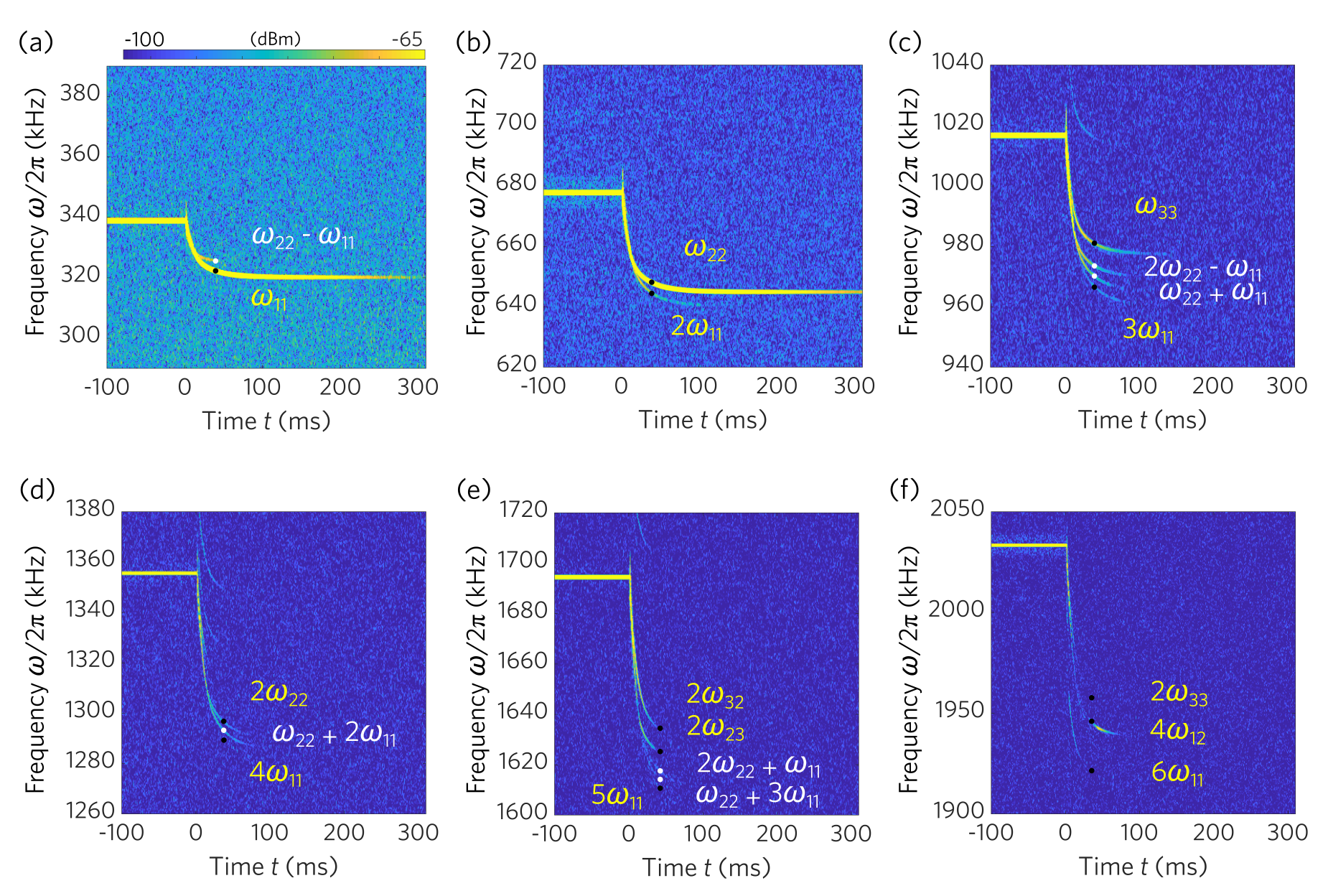}
  \caption[width=\linewidth]{Mode identification of all curves measured in Fig. 3 in the text. The frequencies of the flexural (1,1), (2,2) and (3,3) mode at $t=33~\textrm{ms}$, they can be read from the ring-down spectra: $\omega_{11}/2\pi=322.5~\textrm{kHz}$, $\omega_{22}/2\pi=648.2~\textrm{kHz}$ and $\omega_{33}/2\pi=981.3~\textrm{kHz}$. We only consider the $m = n$ modes for the frequency mixing due to the coupling possibility. We locate all the calculated modulated frequency on the ring-down spectrum by the white dots on figures and labeled as white mode names, for the flexural modes been marked as black dots and labeled as yellow mode names.  From (a) to (f) we labeled all identified flexural modes and mixed modes from 1$\,\omega_d$ up to 6$\,\omega_d$.}
  \label{fig:freq_mix}
\end{figure}

\begin{figure}[htbp]
  \includegraphics[width=0.5\linewidth]{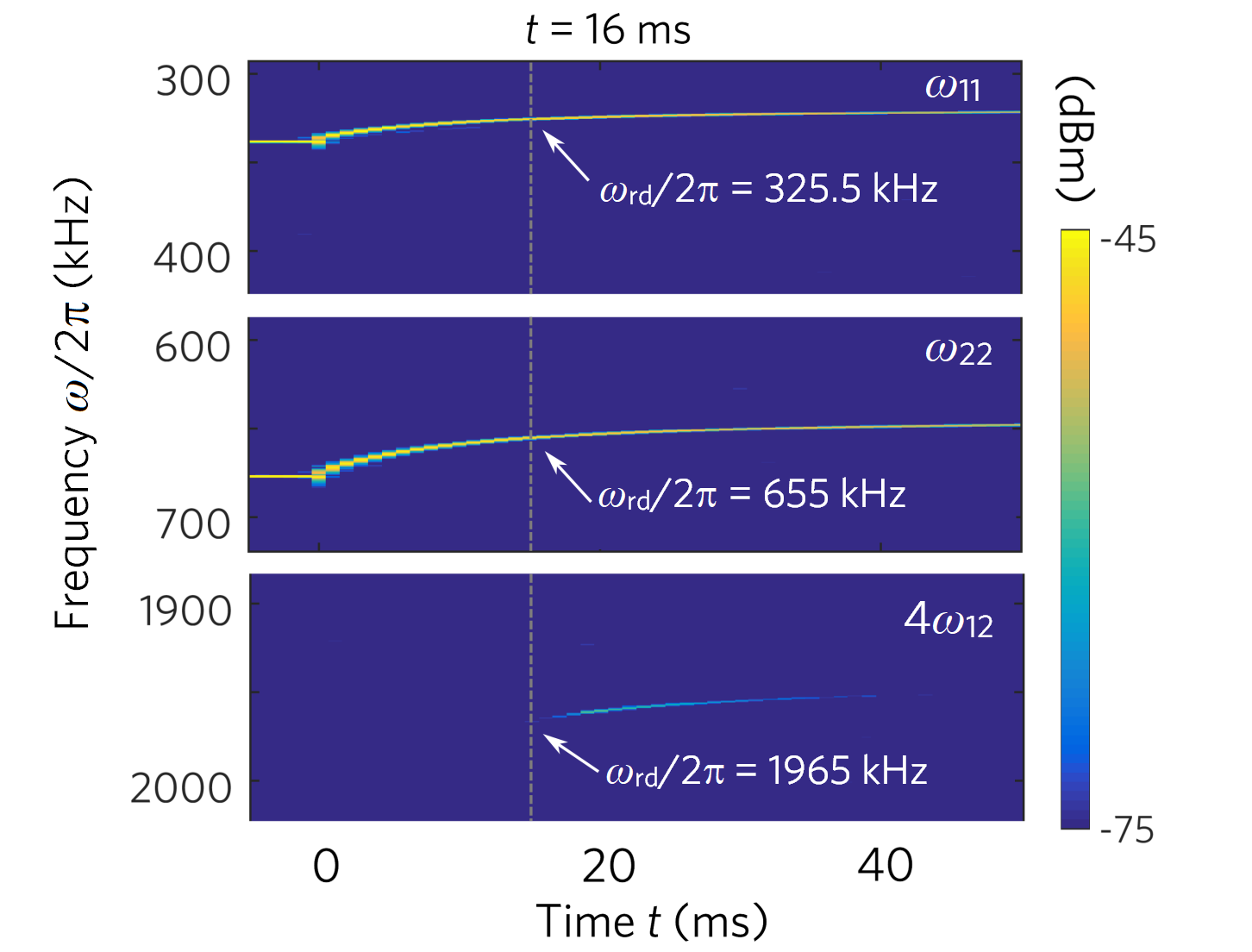}
  \caption[width=\linewidth]{Frequency decay in the strongly driven persistent response regime of the (1,1) mode, selected frequency ranges from the same measurement shown in Fig. 3 in the main text. The indicated values of $\omega_\textrm{rd}$ in each frequency ranges are reading at the time $t$ = 16 ms.}
  \label{fig:Dri_mod_rd}
\end{figure}

\noindent Other traces are produced by frequency mixing between flexural modes and their overtones. We propose that the frequency mixing of a specific mode can be predicted by $\omega_\textrm{mix}$ = $i\,\omega_{11} \pm j\,\omega_{22} \pm k\,\omega_{33} \pm ... \pm z\,\omega_{mn}$, here $i$, $j$, $k$ ... $z$... are integer and we only consider the $m$ = $n$ modes for the frequency mixing due to the coupling possibility. The mixed frequencies have been identified and marked in the Fig. \ref{fig:freq_mix}. The flexural modes and their overtones contribute to the frequency mixing behaviour are marked as black dots and labeled as yellow mode names. To identify the observed mixed modes we calculated the $\omega\textrm{mix}$ by using $\omega_{11}$ and $\omega_{22}$ with the combination index up to 3. We select the $t$ = 33 ms to identify the mode frequencies, ensure that all modes are complete separated during the ring-down process. Here, $\omega_{11}/2\pi$ = 322.5 kHz, $\omega_{22}/2\pi$ = 648.2 kHz at $t$ = 33 ms. We locate all the calculated mixed frequencies (e.g., $\omega_{11}~\pm~\omega_{22}$, 2$\,\omega_{22}~\pm~\omega_{11}$, $\omega_{22}~\pm~2\,\omega_{11}$ and $\omega_{22}~\pm~3\,\omega_{11}$, ect.) on the ring-down spectrum by the white dots on figures show nice consistence with the measured ring-down data, names of mixed modes are labeled as white. \\ 

\noindent In Fig. \ref{fig:Dri_mod_rd}, we present the same data as shown in Fig. 4(a) of the main text. We zoom into the ring-down frequency response around the $\omega_{11}$, $\omega_{22}$, and the 4$\,\omega_{12}$ modes. At \textit{t} = 16 ms after switching off the drive, the 4$\,\omega_{12}$ mode is generated abruptly at $\omega_\textrm{rd}$ = 1965 kHz. As shown in Fig 4(a) in the main text, the $\omega_{33}$, 2$\,\omega_{23/32}$ and 2$\,\omega_{34}$ modes are decaying fast and to the noise floor within the first 16 ms, while the (1,1) and the (2,2) mode decay more slowly and shift their frequencies. At \textit{t} = 16 ms the ring-down frequencies of the modes are 325.5 kHz and 655 kHz, respectively. We noticed that the $\omega_\textrm{rd}$ of 4$\,\omega_{12}$ mode at \textit{t} = 16 ms is exactly three times the actual frequency of the (2,2) mode. Hence, because of the equality 4$\,\omega_{12}$ = 3$\,\omega_{22}$ at this particular moment, we argue that the $4^\textrm{th}$ overtone of the (1,2) mode is generated by energy transfer from the (2,2) mode during the ring-down process. No other integer factor relation between mode eigenfrequencies (up to mode (4,5)) or their overtones (up to $4^\textrm{th}$ overtone) are fulfilled at that time.\\

\begin{figure}[htbp]
  \includegraphics[width=0.5\linewidth]{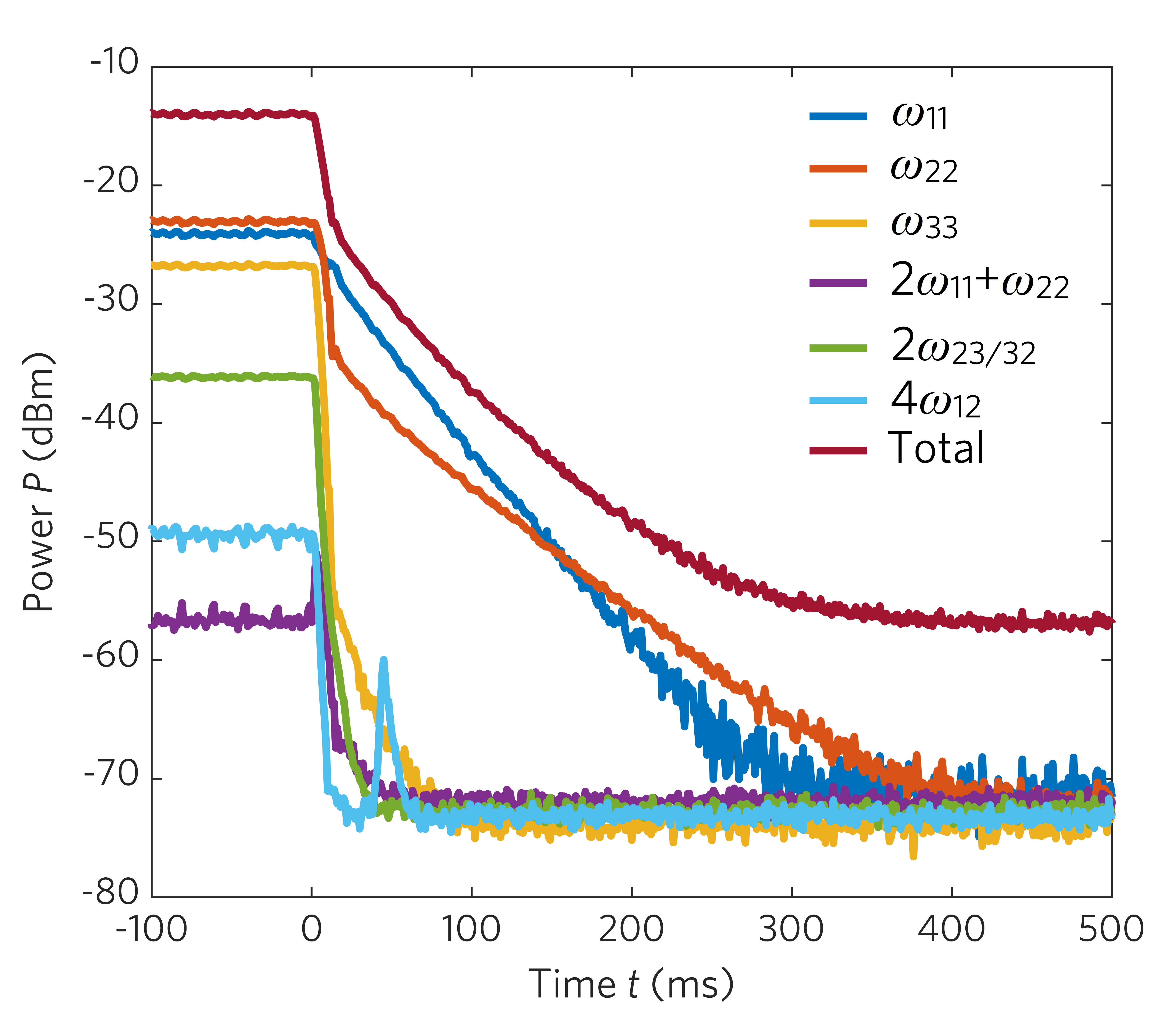}
  \caption[width=\linewidth]{Time resolved, and frequency integrated power spectra of the data shown in main text of Fig. 4(a) for different flexural modes.}
  \label{fig:flex_coup_rd}
\end{figure}
\noindent In Fig. \ref{fig:flex_coup_rd}, the power decay of the resonances in the frequency range from 300 kHz to 2.5 MHz are shown. To interpret the data, one has to keep in mind that the measurement is local, meaning that sudden jumps up or down of the amplitude of particular modes are not violating energy conservation, but just indicate a rearrangement of the spatial distribution of vibration power. At the spot investigated here, the (1,1) mode shows a simple exponential energy decay. The power of the (2,2) mode jumps down immediately after switching off the drive and then also decays exponentially, but slower than the (1,1) mode. For the higher harmonics, the decay behavior is more complex, also involving spatial overtones of the flexural modes \cite{antoni2013nonlinear,polunin2016characterization,yang2019spatial,schuster2009reviews} and the sub-harmonic parametric resonance mechanism discussed above. The decay of the (2,3)/(3,2) modes are exponential with rather small time constant. In the decay of 6$\,\omega_d$, the 4\textsuperscript{th} overtone of the (1,2) mode abruptly jumps up at $\sim$ 30 ms, when the frequency of the (2,2) mode has shifted such that the condition 4$\,\omega_{12}$ = 3$\,\omega_{22}$ is fulfilled. This sudden power increase of 4$\,\omega_{12}$ is reproducible when measuring repeatedly at the same spot, but different mode combinations are detected when measuring at another spot on the membrane as shown in Fig. \ref{fig:Dri_mod_rd}. The common phenomenon is a mutual energy transfer between several flexural modes and their overtones whenever an internal resonance condition is fulfilled during ring-down \cite{guttinger2017energy,chen2017direct}. This behavior is consistent with what we observed by IWLI shown in Fig. 2(a) in the shaded areas in the main text, where at particular driving frequencies internal resonances are met. This observation of an overall redistribution of the excited modes and overtones can be interpreted as an indication for the non-equilibrium property of the persistent response state obtained by the ultra-strong and off-resonant drive.


\section{\label{sec:level8} Theoretical models}
The full theoretical description of the nonlinear coupling between different types of tones and modes and the complex superposition of these internal resonances is quite challenging. 
To explain the experimental observations qualitatively, we focus on the underlying process of each feature and show, solving the theoretical models for three individual cases, that each model explains one of the observations.

We first develop a model for the nonlineraly excited fundamental mode (1,1) of the membrane with eigenfrequency $\omega_{11}$. The goal is to explain qualitatively the experimental observation of the ``persistent response'', namely the two central facts that first, the average vibration amplitude of the membrane is almost flat by increasing the detuning and second, the maximum frequency detuning at which the resonator switches from the high amplitude state to the low amplitude state is much larger with respect to the Duffing model. These two facts can be explained by a consistent theoretical scenario.

In a second step we consider the interaction of the fundamental mode and one specific higher-frequency mode, where we focus on one example for each of two different coupling cases. This is either an indirect parametric nonlinear interaction moderated by the overtones of the fundamental mode, where we study the case of the third overtone of the (1,1) mode with eigenfrequency $3\,\omega_{11}$ which drives indirectly the (2,2) mode with $\omega_{22}\simeq 2\, \omega_{11}$. Or it's a direct parametric nonlinear interaction with a fractional resonance, where we study the interaction of the fundamental mode (1,1) with the (1,2) mode with $\omega_{12}\simeq \frac 32 \, \omega_{11}$.

%
%
%
\subsection{\label{sec:level9} Interaction of the fundamental mode (1,1) with the harmonic modes (2,2), (3,3) and (4,4)}
As explained in the main text, we assume the overtones and the high frequency modes are weakly excited due to nonlinear interaction with the fundamental mode (1,1). 
However, the dynamics of the activated high frequency modes affects the response of the mode $(1,1)$.
To illustrate this idea, we discuss as example a toy model in which we consider the mode $(1,1)$ with eigenfrequency $\omega_{11}\simeq \omega_d$ only coupled to the three modes $(n,n)$ with $n=2,3,4$ and eigenfrequencies $\omega_{nn}\simeq n \, \omega_{11}$ through the potential 
\begin{align}
\sum_{n=2}^{4} V_{ (1\,1 \, |\,  n\,n) }^{(n1)} = \sum_{n=2}^{4}  \lambda_{ (1\,1 \, |\,  n\,n) }^{(n1)} \,\,\, q_{1 1}^n  q_{n n}\, .
\end{align}
Then the dynamical equations read 
\begin{align}
\ddot{q}_{11}=& -\omega_{11}^2q_{11}(t) - 2\Gamma_{11} \dot{q}_{11}(t) - \gamma q_{11}^3(t) + F_d\cos(\omega_d t) - \sum_{n=2}^{4} n~ \lambda_{ (1\,1 \, |\,  n\,n) }^{(n1)} \,\,\, q_{1 1}^{n-1}  q_{n n}\, ,\\
\ddot{q}_{nn}=& -\omega_{nn}^2q_{nn}(t) - 2\Gamma_{nn} \dot{q}_{nn}(t) -  \lambda_{ (1\,1 \, |\,  n\,n) }^{(n1)} \,\,\, q_{1 1}^{n}   \, , \quad  (n \geq 2) \, .
\end{align}
Using the canonical transformations
\begin{align}
u_n(t) =&e^{-in\omega_d t} [q_{nn}(t) - i \dot{q}_{nn}(t)/(n\omega_d)]/2 \, ,\\
u_n^*(t) =&e^{in\omega_d t}~~ [q_{nn}(t) + i \dot{q}_{nn}(t)/(n\omega_d)]/2\, ,
\end{align}
and applying the rotating wave approximation (RWA), we get the equations
\begin{align}
\dot{u}_1(t)=& - \left(	\Gamma_{11} + i\delta\omega_1 - i \frac{3 \gamma}{2\omega_d} |u_1(t)|^2	\right) u_1(t) - i\frac{F}{4\omega_d} +  \sum_{n=2}^{4} \frac{in}{2\omega_d}~ \lambda_{ (1\,1 \, |\,  n\,n) }^{(n1)} \,\,\, u_n(t)(u_1^*(t))^{n-1}\, ,\\
\dot{u}_n(t)=& -\left(\Gamma_{nn} +in\delta\omega_n\right)u_n(t) +i\frac{ \lambda_{ (1\,1 \, |\,  n\,n) }^{(n1)}  }{2n\omega_d} u_1^n(t)    \, , \quad  (n \geq 2) \, , 
\end{align}
with the detuning $\delta\omega_n=\omega_d-\omega_{nn}$. $u(t)$ describes the vibration amplitude in the rotating frame at the driving frequency $\omega_d$. 
We scale the amplitude like  
\begin{align}
u_n(t) =& \sqrt{\frac{2\omega_d \Gamma_{11}}{3\gamma}}z_n(t) \, ,
\end{align}
and introduce new parameters: the scaled detuning, the scaled damping and the scaled force 
\begin{align}
\Omega_n=&\frac{\delta\omega_n}{\Gamma_{11}},~~ \kappa_n=\frac{\Gamma_{nn}}{\Gamma_{11}},~~\beta= \frac{3 F^2 \gamma }{32 \omega_d^3\Gamma_{11}^3} \, .
\end{align}
With these parameters we get the scaled equations
\begin{align}
\dot{z}_1(t)=& - \left(	\Gamma_{11} + i\delta\omega_1 - i \Gamma_{11} |z_1(t)|^2	\right)z_1(t) - i\Gamma_{11}\sqrt{\beta} +  \sum_{n=2}^{4} \frac{in}{2\omega_d}~ \lambda_{ (1\,1 \, |\,  n\,n) }^{(n1)} \,\,\, \left(\sqrt{\frac{2\omega_d \Gamma_{11}}{3\gamma}}\right)^{n-1} z_n(t)(z_1^*(t))^{n-1}\, ,\label{Eq_z1-2}\\
\dot{z}_n(t)=& -\left(\Gamma_{nn} +in\delta\omega_n\right)z_n(t) +i\frac{ \lambda_{ (1\,1 \, |\,  n\,n) }^{(n1)}  }{2n\omega_d} \left(\sqrt{\frac{2\omega_d \Gamma_{11}}{3\gamma}}\right)^{n-1} z_1^n(t)  \, .\label{Eq_zn-2}
\end{align} 

In the steady state, $\dot{z}_n(t)=0$ one can obtain a closed equation from Eq. (\ref{Eq_z1-2}) and (\ref{Eq_zn-2}) for the the stationary solution $\bar{z}_1$. With the abbreviation
\begin{align}
g_n=&\frac{\left(\lambda_{ (1\,1 \, |\,  n\,n) }^{(n1)}\right)^2 }{4\omega_d^2\Gamma_{11}^2}\left(\frac{2\omega_d \Gamma_{11}}{3\gamma}\right)^{n-1}\, ,
\end{align}
we finally get
\begin{align}
-  i\sqrt{\beta} =& \left[	1 + i\Omega_1 - \left(i 	-\frac{g_2 }{\left(\kappa_2+2i\Omega_2\right)}\right)|z_1|^2  +\frac{g_3 }{\left(\kappa_3+3i\Omega_3\right)}~  |z_1|^{4} +\frac{g_4 }{\left(\kappa_4+4i\Omega_4\right)}~  |z_1|^{6}\right]z_1 \nonumber \\
&\simeq
 \left[	1 + i\Omega_1 - i \left(  1+	 \frac{g_2 }{2\Omega_2}\right)|z_1|^2 - i \frac{g_3 }{3\Omega_3}  |z_1|^{4} - i \frac{g_4 }{4\Omega_4}   |z_1|^{6}\right]z_1 
 \, ,
\end{align}
where we used the condition $\kappa_n \ll \Omega_n$.
The latter result corresponds to an effective septic force for the fundamental mode in the RWA.

%
%
\subsection{\label{sec:level10} Theoretical description of fundamental mode as nonlinear resonator with an effective septic potential}

From the previous section, we argue that  
the dynamics of the fundamental mode (1,1) of the membrane with eigenfrequency $\omega_{11}$ can be modeled by an effective septic force. 
Therefore, with the drive $\omega_d\simeq \omega_{11}$ we assume the following equation for the dynamics of the (1,1) mode
\begin{align}
\ddot{q}_{11}(t)=&-\omega_{11}^2q_{11}(t)-2\Gamma_{11}\dot{q}_{11}(t)+F \cos(\omega_d t)-\gamma_1 q^3_{11}(t)-{\mu}q^5_{11}(t)-{\nu}q^7_{11}(t)\, ,\label{Eq_nonlin_fund}
\end{align}
which correpsonds to an effective resonator, driven by a linear force, the Duffing nonlinearity with parameter $\gamma$, a quintic nonlinearity with parameter ${\mu}>0$ and a septic nonlinearity with parameter ${\eta}>0$.

One expects a priori that, for a sufficiently strong drive, the vibration amplitude is large and the nonlinear higher-order terms become more and more important. In particular, the Duffing nonlinearity is not sufficient to explain the dynamics and we need to include these higher-order nonlinearities. The crucial effect of the latter terms is to produce a deflection of the Duffing response curve as a function of the drive frequency. It is also important to note that the inclusion of the quintic (or other higher-order terms) does not imply necessarily that the resonator has more stable solutions than the ones in the Duffing model. As will be shown below, in the regime of parameters that describe the experimental observations of this work, the nonlinear resonator of Eq. (\ref{Eq_nonlin_fund}) has still 3 possible real solutions, out of which two are stable and one unstable, as in the Duffing case. 

Using the canonical transformations
\begin{align}
u(t)=&[q_{11}(t)-{i \dot{q}_{11}(t)}/{\omega_d}  ]e^{-i\omega_d t} \, ,\\
u^*(t)=&[q_{11}(t)+{i \dot{q}_{11} (t)}/{\omega_d} ]e^{i\omega_d t} \,  ,
\end{align}
and applying the RWA we get the equation
\begin{align}
\dot{u}(t)=& \left[-i\delta\omega_d - \Gamma_{11} +i \frac{3 \gamma}{8\omega_d}|u(t)|^2 +i\frac{5{\mu}}{16\omega_d} |u(t)|^4 +i\frac{35 {\nu}}{128\omega_d} |u(t)|^6 \right]u(t)- i\frac{F}{2\omega_d}  \, ,
 \end{align}
 with the detuning $\delta\omega_d = \omega_d-\omega_{11}$. The variable $u(t)$ describes the vibration amplitude in the rotating frame at the driving frequency $\omega_d$. 
We scale the amplitude like
\begin{align}
u(t)=&\sqrt{\frac{8\omega_d\Gamma_{11}}{3\gamma}} z(t) \, ,
\end{align}
and introduce the new parameters: the scaled detuning, the scaled quintic and septic nonlinearity parameters and the scaled force, provided by an external drive,
 \begin{align}
\Omega=\frac{\delta\omega_d}{\Gamma_{11}} \, ,
~~~~~ \tilde{\mu}=\frac{20}{9} {\mu} \frac{\omega_d\Gamma_{11}}{\gamma^2} \, ,
~~~~~ \tilde{\nu}=\frac{140}{27} {\nu} \frac{\omega_d^2\Gamma_{11}^2}{\gamma^3} \, ,
~~~~~ \beta=\frac{3\gamma F^2}{32 \omega_d^3\Gamma_{11}^3} \, .
 \end{align}
With these we get the short equation for the scaled amplitude in the RWA
\begin{align}
\frac{1}{\Gamma_{11}} \dot{z}(t)=& \left(	-i\Omega - 1 +i|z(t)|^2 +i\tilde{\mu}|z(t)|^4+i\tilde{\nu}|z(t)|^6	\right)z(t)-i\sqrt{\beta} \, .
\end{align}
In the steady state, $\dot{z}(t)=0$, the stationary solution $\bar{z}$ is given by the equation
\begin{align}
\beta=&\left[1+\left(\Omega-|\bar{z}|^2-\tilde{\mu}|\bar{z}|^4 - \tilde{\nu} |\bar{z}|^6 \right)\right]|\bar{z}|^2 \, . \label{Eq_steady-state-scale-2}
\end{align}
 If the external driving force is fixed, the parameter $\beta$ is fixed. One can then plot the solutions of Eq. (\ref{Eq_steady-state-scale-2}) as a function of the scaled detuning $\Omega$. With the scaling proposed above, the solutions then depend only on two parameters, namely $\tilde{\nu}$ and $\tilde{\mu}$, as evident from Eq. (\ref{Eq_steady-state-scale-2}). This is shown in Fig. 1 (b) of the main text.
The Duffing resonator is recovered if $\mu=\nu=0$. One can then compare the solutions of the RWA, so Eq. (\ref{Eq_steady-state-scale-2}), for the amplitude with the full numerical simulation of a version of Eq. (\ref{Eq_nonlin_fund}) that is scaled in the same way. Deviations between the full numerical solution and the RWA solution appear at large detuning beyond the validity of the RWA which is given by
\begin{align}
|\omega_d-\omega_{11}|\ll \omega_d,~~~|\omega_d-\omega_{11}|\ll \omega_{11} \, .
\end{align}
 The RWA breaks down when the detuning is comparable to $\omega_{11}$. 
 
 Solving Eq. (\ref{Eq_steady-state-scale-2}) for the detuning $\Omega$, we assume that we have either 1 or 3 real solutions for $|\bar{z}|^2$.
Then the condition  $\beta \, /\, |\bar{z}|^2-1=0$ set the maximum amplitude $|z_{\text{max}}|^2=\beta$ leads to the maximum detuning of
 \begin{align}
\Omega_{\text{max}} =&|\bar{z}|^2_{\text{max}}+\tilde{\mu} (|\bar{z}|^2_{\text{max}})^2+ \tilde{\nu} (|\bar{z}|^2_{\text{max}})^3
	                = \beta+\tilde{\mu} \beta^2 + \tilde{\nu}\beta ^3 \, .
 \end{align}
Without the scaling, the maximum amplitude and maximum detuning become
 \begin{align}
|u_{\text{max}}|^2=& \frac{F^2}{4\omega_d^2 \Gamma_{11}^2} \, ,\\
\frac{\delta\omega_d}{\Gamma_{11}} =& \frac{3\gamma F^2}{32\omega_d^3\Gamma_{11}^3} + \frac{5{\mu}F^4}{256\omega_d^5\Gamma_{11}^5} + \frac{35{\nu}F^6}{2^{13}\omega_d^7\Gamma_{11}^7} \, .
 \end{align}
As it is possible to see in Fig. 1 (b) of the main text, the curve of the amplitude vs. detuning becomes flatter by adding nonlinear self-interaction terms in the equation. This explains qualitatively the observed saturation in the measurements.\\

%
%
%
\subsection{\label{sec:level11} Indirect parametric nonlinear interaction activated by the third overtone of the fundamental mode}

For the first coupling mechanism, the indirect parametric nonlinear interaction activated by the overtones of the fundamental mode, 
we study the specific case of the third overtone of the (1,1) mode with eigenfrequency $3 \, \omega_{11}$ which drives indirectly 
the (2,2) mode with $\omega_{22}\simeq 2  \, \omega_{11}$. 
We set the following coupled equations for the amplitudes $q_{11}(t)$ and $q_{22}$(t)
 \begin{align}
 \ddot{q}_{11}(t)=&-\omega_{11}^2q_{11}(t)-2\Gamma_{11}\dot{q}_{11}(t)+F \cos(\omega_d t)-\gamma_1(t)q_{11}^3(t)-\lambda q_{11}(t) q_{22}^2(t) \, ,\label{Eq_1:2_q1}\\
 \ddot{q}_{22}(t)=&-\omega_{22}^2q_{22}(t)-2\Gamma_{22}\dot{q}_{22}(t)-\gamma_2q_{22}^3(t)-\lambda q_{11}^2(t)q_{22}(t) \, .\label{Eq_1:2_q2}
 \end{align}
  Both effective resonators of Eq. (\ref{Eq_1:2_q1}) and (\ref{Eq_1:2_q2}) are characterized by a Duffing nonlinearity with strength $\gamma_1$ or $\gamma_2$, the eigenfrequencies $\omega_{11}$ and $\omega_{22}$, and a damping constant $\Gamma_{11}$ or $\Gamma_{22}$. The first resonator is linearly driven by the force $F$. The nonlinear interaction of the two resonators is described by the potential $V^{(22)}_{(11|22)}= 1/2 \, \lambda^{(22)}_{(11|22)} \, q_{11}^2 \,  q_{22}^2$ with interaction strength $\lambda\equiv \lambda^{(22)}_{(11|22)}$.
We use the following anzats to include also the presence of the overtone of the fundamental mode
 \begin{align}
 q_{11}(t)=& \frac 12 [u_1(t) e^{i\omega_dt}  + u_1^*(t)  e^{-i\omega_dt}]  + \frac 12  [u_3(t) e^{3i\omega_d t}+u_3^*(t) e^{-3i\omega_d t}]\, ,\\
 q_{22}(t)=& \frac 12 [v(t) e^{2i\omega_dt} +v^*(t) e^{-2i\omega_dt}]\, ,
 \end{align}
and apply the RWA we get then three coupled equations 
%
%
%
\begin{align}
0=&\left(-i\delta\omega_1-\Gamma_{11}+i\frac{3\gamma_1}{8\omega_d}|u_1(t)|^2+i\frac{3\gamma_1}{4\omega_d} |u_3(t)|^2+i\frac{\lambda}{4\omega_d}|v^2(t)|\right)u_1 -i \frac{F}{2\omega_d} +i\frac{3\gamma_1}{8\omega_d} (u_1^*(t))^2u_3(t) 	 +i\frac{\lambda}{8\omega_d} u_3^*(t)v^2(t) \, ,\\
0
	=& \left(4-\frac{3\gamma_1}{8\omega_d^2} |u_3(t)|^2-\frac{3\gamma_1}{4\omega_d^2}|u_1(t)|^2-\frac{\lambda}{4\omega_d^2}|v|^2\right)u_3(t)  -\frac{\gamma_1}{8\omega_d^2}	u_1^3(t)   -\frac{\lambda}{8\omega_d^2}u_1^*(t)v^2(t) \, ,\\
0=& \left(-i\delta\omega_2-\Gamma_{22}+i\frac{3\gamma_2}{16\omega_d}|v^2(t)| +i \frac{\lambda}{8\omega_d} \left(|u_1(t)|^2+|u_3(t)|^2\right)\right)v(t)  +i \frac{\lambda}{8\omega_d} u_1(t)u_3(t)v^*(t) \, ,
\end{align} 
with the detunings $\delta\omega_1=\omega_d-\omega_{11}$ and $\delta\omega_2=2\omega_d-\omega_{12}$.
We use the following scaling for the amplitudes
 \begin{align}
 z_1(t)=& \sqrt{\frac{3\gamma_1}{8\omega_d \Gamma_{11}}}u_1(t) \, ,~~
 z_3(t)= \sqrt{\frac{3\gamma_1}{8\omega_d \Gamma_{11}}}u_3(t) \, ,~~
 w(t)= \sqrt{\frac{3\gamma_2}{8\omega_d \Gamma_{22}}}v(t) \, ,
 \end{align}
 and the parameters for the difference between the resonance frequencies scaled by $\Gamma_{11}$, the scaled force, the detuning of the first resonator scaled by $\Gamma_{11}$ and the detuning of the second resonator scaled by $\Gamma_{22}$, a coupling parameter and an asymmetry factor 
 \begin{align}
 \Delta=\frac{\Delta\omega}{\Gamma_{11}} \, ,~~~ 
\sqrt{\beta}=\sqrt{\frac{3\gamma_1F^2}{32\omega_d^3\Gamma_{11}^3}} \, ,~~~
 \Omega_1=\frac{\delta\omega_1}{\Gamma_{11}} \, ,~~~
 \Omega_2= \frac{\delta\omega_2}{\Gamma_{22}}=\frac{\Gamma_{11}}{\Gamma_{22}}(2\Omega_1-\Delta) \, ,~~~
 g=\frac{\lambda}{\sqrt{\gamma_1\gamma_2}} \, ,~~~
 \alpha=\sqrt{\frac{\gamma_1}{\gamma_2}}\left(\frac{\Gamma_{22}}{\Gamma_{11}}\right) \, .
 \end{align}

The scaled coupled equations read as follows: 

  \begin{align}
   &0=\left(-1-i\Omega_1+ i |z_1(t)|^2 + 2 i |z_3(t)|^2 + i\frac{2}{3} g \alpha	|w(t)|^2	\right) z_1(t) -i\sqrt{\beta}	 + i(z_1^*(t))^2 z_3(t)+i \frac{g}{3}\alpha z_3^*(t) w^2(t) \\
   &0=	\left(	4-\frac{\Gamma_{11}}{\omega_d}|z_3(t)|^2-2\frac{\Gamma_{11}}{\omega_d}|z_1(t)|^2-\frac 23 g\alpha\frac{\Gamma_{11}}{\omega_d}|w(t)|^2 \right)z_3(t)-\frac{\Gamma_{11}}{3\omega_d}z_1^3(t)-\frac 13 \frac{\Gamma_{11}}{\omega_d} g\alpha z_1^*(t)w^2(t) 	 \\
   &0=\left(	-1-i\frac{\Gamma_{11}}{\Gamma_{22}}(2\Omega_1-\Delta)	+\frac i2  |w(t)|^2 +\frac i3  \frac{g}{\alpha} \left(|z_1(t)|^2+|z_3(t)|^2\right)		\right) w(t) + \frac i3 \frac{g}{\alpha} z_1(t) z_3 w^*(t)
   \end{align}
 For $w(t) \ne 0$ the last equation can be transformed to a direct expression for $|w(t)|$ depending on $z_1(t)$ and $z_3(t)$. Assuming $\frac{\Gamma_{11}}{\omega_d} |z_1|^2\ll 1,~~ \frac{\Gamma_{11}}{\omega_d} |z_3|^2\ll 1,~~ \frac{\Gamma_{11}}{\omega_d} |w|^2\ll 1$ and $\alpha\ll 1$ we can simplify the equations to
  \begin{align}
    &0=\left(-1-i\Omega_1+ i |z_1|^2 + 2 i |z_3|^2 + i\frac{2}{3} g \alpha	|w|^2		\right) z_1 -i\sqrt{\beta}  + i(z_1^*)^2 z_3\label{Eq_1-alpha} \\
  &z_3=\frac{\Gamma_{11}}{12\omega_d} z_1^3\label{Eq_2-alpha}\\
 & |w|^2=2 \frac{\Gamma_{11}}{\Gamma_{22}} (2\Omega_1-\Delta) -\frac{2g}{3\alpha} \left(	|z_1|^2+|z_3|^2	\right) \pm 2 \sqrt{\frac 19 \left(\frac{g}{\alpha}\right)^2 |z_1|^2|z_3|^2-1 } \label{Eq_3-alpha}
    \end{align}
Inserting Eq. (\ref{Eq_2-alpha}) in Eq. (\ref{Eq_1-alpha}) yields a closed equation for $|z_1|$.
Inserting Eq. (\ref{Eq_2-alpha}) in Eq. (\ref{Eq_3-alpha}) we obtain an equation for $|w|$ depending only on $|z_1|$ that has been determined by the closed equation before.
We solved this problem   numerically. The result is shown in the right panel of Fig. 2(b) in the main text.
Notice that  the result for $z_1$ is having an effective high-order nonlinearities.

\subsection{\label{sec:level12} Direct parametric nonlinear interaction caused by fractional resonance}
For the direct parametric nonlinear interaction with higher flexural modes ($m \ne n$), so caused by a fractional resonance, we consider the interaction of the fundamental mode (1,1) with eigenfrequency $\omega_{11}$, with the (1,2) mode with frequency $\omega_{12}\approx 3/2  \, \omega_{11}$. To this end, we treat both modes as individual resonators and consider a nonlinear coupling between both. 
We set the following coupled equations for the amplitudes $q_{11}(t)$ and $q_{12}(t)$:
\begin{align}
\ddot{q}_{11}(t) =& -\omega_{11}^2 q_1(t) - 2\Gamma_{11}\dot{q}_{11}(t)+F\cos(\omega_d t)-\gamma_1q_{11}^3(t)-\frac 32 \lambda  q_{11}^2(t)q_{12}^2(t) \, ,\label{Eq_2:3_q1}\\
\ddot{q}_{12} (t)=& -\omega_{12}^2 q_{12}(t)- 2\Gamma_{12}\dot{q}_{12}(t)-\gamma_2q_{12}^3(t)- \lambda q_{11}^3(t)q_{12}(t) \, .\label{Eq_2:3_q2}
 \end{align}
 Both effective resonators of Eq. (\ref{Eq_2:3_q1}) and (\ref{Eq_2:3_q2}) are characterized by a Duffing nonlinearity with strength $\gamma_1$ or $\gamma_2$, the eigenfrequencies $\omega_{11}$ and $\omega_{12}$, and a damping constant $\Gamma_{11}$ or $\Gamma_{12}$. The first resonator is linearly driven by the force $F$. An essential ingredient is the nonlinear interaction of the two resonators, described by the potential $V^{(32)}_{(11|12)}=\frac 12  \, \lambda^{(32)}_{(11|12)}  \, q_{11}^3  \, q_{12}^2$ with interaction strength $\lambda\equiv \lambda^{(32)}_{(11|12)}$.
Using the canonical transformations
\begin{align}
u_1(t)=& [q_{11}(t)-{i \dot{q}_{11}(t)}/{\omega_d}]e^{-i\omega_dt},~~~
u_1(t)^*= [q_{11}(t)+{i \dot{q}_{11}(t)}/{\omega_d}]e^{i\omega_dt} \, ,\\
u_2(t)=& [q_{12}(t)-{i2\dot{q}_{12}(t)}/{(3\omega_d)}]e^{-i\frac 32 \omega_dt},~~~
u_2(t)^*= [q_{12}(t)+{i2\dot{q}_{12}(t)}/{(3\omega_d)}]e^{i\frac 32 \omega_dt} \, ,
\end{align}
and applying the RWA we obtain two coupled equations 
\begin{align}
\dot{u}_1(t)= & \left(-i\delta\omega_1-\Gamma_{11}+i\frac{3\gamma_1}{8\omega_d}|u_1(t)|^2\right)u_1(t)-i\frac{F}{2\omega_d}+i\frac{3}{32}\frac{\lambda}{\omega_d}\left(u_1^*(t)\right)^2u_2^2(t) \, ,\\
\dot{u}_2(t)=& \left(-i\frac{3}{2}\delta\omega_2-\Gamma_{12}+i\frac{\gamma_2}{4 \omega_d}|u_2(t)|^2\right)u_2(t)+i\frac{\lambda}{24\omega_d}u_1^3(t) u_2^*(t) \, ,
\end{align}
with the detunings $\delta\omega_1=\omega_d-\omega_{11}$ and $\delta\omega_2=\omega_d- 2/3 \, \omega_{12}$.  The variables $u_{1}(t)$ and $u_{2}(t)$ describe the vibration amplitude of the two effective resonators.
We use the following scaling for the two amplitudes
 \begin{align}
 u_1(t)=& \sqrt{\frac{8\omega_d\Gamma_{11}}{3\gamma_1}} z_1(t),~~~  u_2(t)= \sqrt{\frac{4\omega_d\Gamma_{12}}{\gamma_2}}z_2(t) \, ,
 \end{align}
 and directly introduce the parameter $\sqrt{\beta}$ for the scaled force and the abbreviations $g$ and $h$
\begin{align}
\beta= \frac{3\gamma_1F^2}{32\omega_d^3\Gamma_{11}^3} \, ,~~~
g=\frac{27}{8} \left(\frac{\Gamma_{12}}{\Gamma_{11}}\right)^2 \left(\frac{\gamma_1}{\gamma_2}\right)h \, ,~~~
h= \frac{\lambda}{9}\left(\frac{\Gamma_{11}}{\Gamma_{12}}\right)\sqrt{\frac{8\omega_d\Gamma_{11}}{3\gamma_1^3}} \, .
\end{align}
%
In the steady state $\dot{z}_1=\dot{z}_2=0$, the stationary solutions $\bar{z}_{1}$ and $\bar{z}_{2}$ are given by the two coupled equations  
\begin{align}
0=&\left(-i\frac{\delta\omega_1}{\Gamma_{11}}-1+i|\bar{z}_1|^2\right)\bar{z}_1-i\sqrt{\beta}+ig(\bar{z}_1^*)^2\bar{z}_2^2 \, ,\label{Eq_ss-1}\\
0=&\left(-i\frac{\frac 32 \delta\omega_2}{\Gamma_{12}}-1+i|\bar{z}_2|^2\right)\bar{z}_2+ih\bar{z}_1^3(t)\bar{z}_2^* \, .\label{Eq_ss-2}
\end{align}
Transforming Eq. (\ref{Eq_ss-2}) and taking the square modulus of both sides yields an expression for $|\bar{z}_2|\ne 0$ depending only on $|\bar{z}_1|$
\begin{align}
|\bar{z}_2|^2=\frac{3\delta\omega_2}{2\Gamma_{12}}+\sqrt{h^2|\bar{z}_1|^6-1} \, .\label{Eq_ss_2.2}
\end{align}
To transform Eq. (\ref{Eq_ss-1}) we focus first on the phase via $\bar{z}_i=|\bar{z}_i|e^{i\Theta_i}$. 
With Eq. (\ref{Eq_ss_2.2}) and the phase representation for the two amplitudes we can approximate from Eq. (\ref{Eq_ss-1}) the phase relation $\Theta_2=3/2 \Theta_1$ in the long amplitude limit. With this phase relation Eq. (\ref{Eq_ss-1}) can be transformed into a phase-independent equation
\begin{align}
\beta=& \left[1+\left(\frac{\delta\omega_1}{\Gamma_{11}}-|\bar{z}_1|^2-g|\bar{z}_1||\bar{z}_2|^2\right)	\right]|\bar{z}_1|^2 \, .
\end{align}
Replacing $|\bar{z}_2|$ by Eq. (\ref{Eq_ss_2.2}) we get a closed equation for $|\bar{z}_1|$ and solve the problem numerically for
\begin{align}
&\left[1+\left(\frac{\delta\omega_1}{\Gamma_{11}}-|\bar{z}_1|^2-g|\bar{z}_1|\left[\frac{3\delta\omega_2}{2\Gamma_{12}}+\sqrt{h^2|\bar{z}_1|^6-1}\right]\right)^2\right]|\bar{z}_1|^2=\beta \, .
\end{align}
The result is shown in the left panel of Fig. 2(b) of the main text.

\bibliographystyle{apsrev4-1}
\bibliography{references}